# A Theory of Complexity, Condition and Roundoff

Felipe Cucker*
Department of Mathematics
City University of Hong Kong
HONG KONG
e-mail: macucker@cityu.edu.hk

**Abstract**

We develop a theory of complexity for numerical computations that takes into account the condition of the input data and allows for roundoff in the computations. We follow the lines of the theory developed by Blum, Shub, and Smale for computations over $\mathbb{R}$ (which in turn followed those of the classical, discrete, complexity theory as laid down by Cook, Karp, and Levin among others). In particular, we focus on complexity classes of decision problems and paramount among them, on appropriate versions of the classes $\mathsf{P}$, $\mathsf{NP}$ and $\mathsf{EXP}$ of polynomial, nondeterministic polynomial, and exponential time, respectively. We prove some basic relationships between these complexity classes and exhibit natural NP-complete problems.

## Contents

*Partially funded by a GRF grant from the Research Grants Council of the Hong Kong SAR (project number CityU 100810).

# 1 Introduction

## 1.1 Background

A quarter of a century ago —give or take a month or two— Lenore Blum, Mike Shub, and Steve Smale published an article [6] developping a theory of complexity over the real numbers. The two previous decades had witnessed a spectacular development of the foundations of discrete computations and the declared purpose of [6] was to lay down the grounds for a similar development for numerical computations. To understand the nature of this goal it will be useful to give an overview of the ideas involved.

The design of computer software (operating systems, compilers, text editors) that accompanied the spread of digital computers brought an interest in the understanding of the cost of algorithmic solutions for a large number of combinatorial problems (searching, sorting, pattern matching). This interest took two forms: the analysis of specific algorithms and the search of inherent lower bounds for specific problems. The former would allow to compare the efficiency of different algorithms whereas the latter would allow to compare any algorithm's cost with current lower bounds and, in some cases, prove optimality.

On the other extreme of optimality results, a number of problems exhibited a large gap between the cost of their best algorithmic solutions and their provable lower bounds. To understand this gap, and to eventually decide which of the bounds was off the mark, it was beside the point to use a cost measure that would be too fine. Instead, emphasis was soon made on *polynomial time* as opposed to *exponential time*, the former broadly meaning "tractable" and the latter "intractable" [14, 28]. Furthermore, technical reasons allowed to focus on decision problems (questions with a `Yes/No` answer) and this gave rise to the classes $\mathsf{P}$ and $\mathsf{EXP}$ of such problems solvable in polynomial and exponential time, respectively. The existence of problems in $\mathsf{EXP} \setminus \mathsf{P}$ was soon established [32] but these problems were somehow artificial and had no relevance besides helping to establish that $\mathsf{P} \neq \mathsf{EXP}$. For many problems of interest, the question of whether a polynomial time algorithm could be devised (or, instead, a superpolynomial complexity lower bound proved) remained open.

A new insight developed in the early 70s that had a lasting impact on theoretical computer science. There was a surge of interest on a subclass of $\mathsf{EXP}$ consisting in searching problems for which a candidate solution can easily (i.e., in $\mathsf{P}$) be verified to be (or not) a true solution to the search. This class was given the name $\mathsf{NP}$ (from *nondeterministic polynomial time*) and satisfied the inclusions $\mathsf{P} \subset \mathsf{NP} \subset \mathsf{EXP}$. Then, Steve Cook [15] and Leonid Levin [41] independently proved that the problem $\mathsf{SAT}$ of deciding whether a Boolean formula had a satisfying assignment (i.e., whether its variables can be given values in $\{\mathtt{True}, \mathtt{False}\}$ such that the formula evaluates to `True`) had the following properties:

(i) $\mathsf{SAT}$ belongs to $\mathsf{NP}$, and

(ii) if $\mathsf{SAT}$ belongs to $\mathsf{P}$ then $\mathsf{P} = \mathsf{NP}$.

Shortly after, Richard Karp [37] showed that 21 problems coming from diverse areas of discrete computation shared these two properties as well, and it was a matter of a few years to have literally thousands of them. These problems are said to be $\mathsf{NP}$-*complete*. And, as it happens, the membership in point (ii) above remains open for all of them. That is, it is not known whether any of them is in $\mathsf{P}$, or equivalently, whether $\mathsf{P} = \mathsf{NP}$.

The lanscape drawn by these results is frustrating. We can prove exponential lower bounds for some natural problems but these are few. On the other hand, we can prove $\mathsf{NP}$-completeness for a large number of problems but cannot deduce superpolynomial lower bounds from these completeness results since we do not know whether $\mathsf{P} = \mathsf{NP}$. One can therefore understand that the truth of this equality became the most important open problem in theoretical computer science, and even a paramount one for mathematicians [16, 57].

The $\mathsf{P} = \mathsf{NP}$ question along with everything involved in it (notably, a formal machine model upon which a notion of cost can be defined) revealed a gap between the theoretical foundations of discrete computations in the early 80s and those of numerical computations. It is this gap (indeed, the desire to fill it) what motivated

Blum, Shub and Smale. Among other results, their paper [6] defined a formal computational model over the real numbers, associated a natural cost measure to computations in this model, and used this cost measure to define complexity classes $\mathrm{P}_{\mathbb{R}}$, $\mathrm{NP}_{\mathbb{R}}$, and $\mathrm{EXP}_{\mathbb{R}}$ satisfying the following properties (all of them, mimicking known properties of their discrete counterparts):

(i) the classes $\mathrm{P}_{\mathbb{R}}$ and $\mathrm{EXP}_{\mathbb{R}}$ are closed by complements (i.e., if a problem $S$ is in the class, so is the problem obtained by exchanging `Yes` and `No` answers),

(ii) $\mathrm{P}_{\mathbb{R}} \subset \mathrm{NP}_{\mathbb{R}} \subset \mathrm{EXP}_{\mathbb{R}}$, and

(iii) the class $\mathrm{NP}_{\mathbb{R}}$ has natural complete problems.

The $\mathrm{NP}_{\mathbb{R}}$-complete problem exhibited in [6] is the following: given a polynomial $f \in \mathbb{R}[X_1, \dots, X_n]$ of degree at most 4, does there exists $\xi \in \mathbb{R}^n$ such that $f(\xi) = 0$? Unlike the situation in the discrete setting, however, there was no avalanche of $\mathrm{NP}_{\mathbb{R}}$-complete problems after the publication of [6]. We won't delve into the reasons of this contrast (the interested reader may find a possible cause in [8]). Also, we note here that the inclusion $\mathrm{NP}_{\mathbb{R}} \subset \mathrm{EXP}_{\mathbb{R}}$ was not proved in [6] and that it is certainly non-trivial (see, e.g., [33, 45]). It is, in addition, strict (i.e., $\mathrm{NP}_{\mathbb{R}} \neq \mathrm{EXP}_{\mathbb{R}}$, see [17]), a separation that in the discrete setting remains conjectural as of today.

The ideas in [6] fusioned algebraic complexity theory and structural complexity and, simultaneously, built a bridge between theory of computation and numerical analysis. Its influence after a quarter of century —give or take a month or two— can hardly be overestimated.

The above notwithstanding, Blum, Shub and Smale were aware of at least two aspects left out of their exposition. Firtsly, the consideration of roundoff errors and their effect on computation. Secondly, the complexity of iterative methods. Both issues are related to the notion of *condition* (the analyses of both are expressed in terms of a *condition number*) and are the warp and woof of numerical analysis. Actually, the last section of [6] is devoted to open problems, the last of which reads

> Finally, to bring machines over $\mathbb{R}$ closer to the subject of numerical analysis, it would be useful to incorporate round-off error, condition numbers and approximate solutions into our development.

Our only agenda here is to pursue this proposal.

## 1.2 Main results and structure of the exposition

In this paper we extend the notion of *decision problem* to include the condition of an input. Besides the *length* of such an input there is a natural notion of *size* that naturally takes into account its condition. Also, for any finite-precision computation we define a notion of *cost* that accommodates precision requirements.

Endowed with these basic notions the first goal is to define a version of the class $\mathsf{P}$ in the context of finite precision. Our version is the class $\mathrm{P}_{\mathsf{ro}}$ of problems decidable with *roundoff polynomial cost.* This is a very general class that captures, we believe, the features and uncertainties of finite-precision computations. The complexity class $\mathrm{EXP}_{\mathsf{ro}}$ is similarly defined. Both $\mathrm{P}_{\mathsf{ro}}$ and $\mathrm{EXP}_{\mathsf{ro}}$ are closed by complements. In other words, property (i) above holds in this setting as well.

Thinking about nondeterminism leads to the issue of which candidate solutions $y$ are taken into account for a given problem. In this paper we consider two possible answers to this question according to whether the magnitude of such a $y$ (roughly speaking, the lengths of the exponents in a floating-point representation of the components $y_i$) is to be polynomially bounded on the size of the problem's data. We thus obtain the classes of *bounded* and *unrestricted* nondeterministic polynomial time $\mathrm{NP}^{\mathsf{B}}_{\mathsf{ro}}$ and $\mathrm{NP}^{\mathsf{U}}_{\mathsf{ro}}$, respectively. For both of these two classes we show that a version of the *Circuit feasibility problem* is complete (under a notion of reduction that is appropriate to our finite-precision context). The question to be decided is the same in both cases but the condition numbers are different.

For these classes we therefore show the following:

(ii') $\mathrm{P}_{\mathsf{ro}} \subset \mathrm{NP}^{\mathsf{B}}_{\mathsf{ro}} \subset \mathrm{NP}^{\mathsf{U}}_{\mathsf{ro}}$, and $\mathrm{NP}^{\mathsf{B}}_{\mathsf{ro}} \subset \mathrm{EXP}_{\mathsf{ro}}$, and

(iii') the classes $\mathrm{NP}^{\mathsf{B}}_{\mathsf{ro}}$ and $\mathrm{NP}^{\mathsf{U}}_{\mathsf{ro}}$ have natural complete problems.

Furthermore, we show that the inclusion $\mathrm{NP}^{\mathsf{B}}_{\mathsf{ro}} \subset \mathrm{EXP}_{\mathsf{ro}}$ is strict, just as the inclusion $\mathrm{NP}_{\mathbb{R}} \subset \mathrm{EXP}_{\mathbb{R}}$ is, but the proof now relies on bounds for precision requirements (which we show in the more general form of a hierarchy theorem). The inclusion $\mathrm{NP}^{\mathsf{U}}_{\mathsf{ro}} \subset \mathrm{EXP}_{\mathsf{ro}}$ remains open; we conjecture it holds true.

The class $\mathrm{P}_{\mathsf{ro}}$ is, as we said, very general. A glance at the literature shows the existence of two, more down-to-earth, subclasses of $\mathrm{P}_{\mathsf{ro}}$ which we will denote by $\mathrm{P}_{\mathsf{dir}}$ and $\mathrm{P}_{\mathsf{iter}}$. In the first (which roughly coincides with the so called *direct* algorithms) the running time of the algorithm does not depend on the precision needed. An error-free execution provides the right answer. In the second, iteration is of the essence and machine precision can be adjusted during the execution. A fundamental feature now is that the algorithm's outputs are guaranteed correct. The corresponding classes for exponential cost naturally follow.

A way in which the class $\mathrm{P}_{\mathsf{ro}}$ becomes "closer to numerical analysis" (than the class $\mathrm{P}_{\mathbb{R}}$ introduced in [6]) is that subsets of $\mathbb{R}^n$ decided in $\mathrm{P}_{\mathsf{ro}}$ no longer need to have semialgebraic boundaries. An example is the set $\{(x, y) \mid x \in (0, 1) \text{ and } y \leq e^x\}$ which belongs to $\mathrm{P}_{\mathsf{ro}}$ (when endowed with a natural condition nuber).

Four further classes for nondeterministic polynomial cost can be naturally defined. Yet, in our exposition, we won't elaborate on $\mathrm{NP}^{\mathsf{U}}_{\mathsf{iter}}$ or $\mathrm{NP}^{\mathsf{B}}_{\mathsf{iter}}$, and we will merely briefly discuss on the classes $\mathrm{NP}^{\mathsf{U}}_{\mathsf{dir}}$ and $\mathrm{NP}^{\mathsf{B}}_{\mathsf{dir}}$, for which a completeness result is shown. A diagram of the main classes studied in this paper, with arrows indicating inclusions, follows.

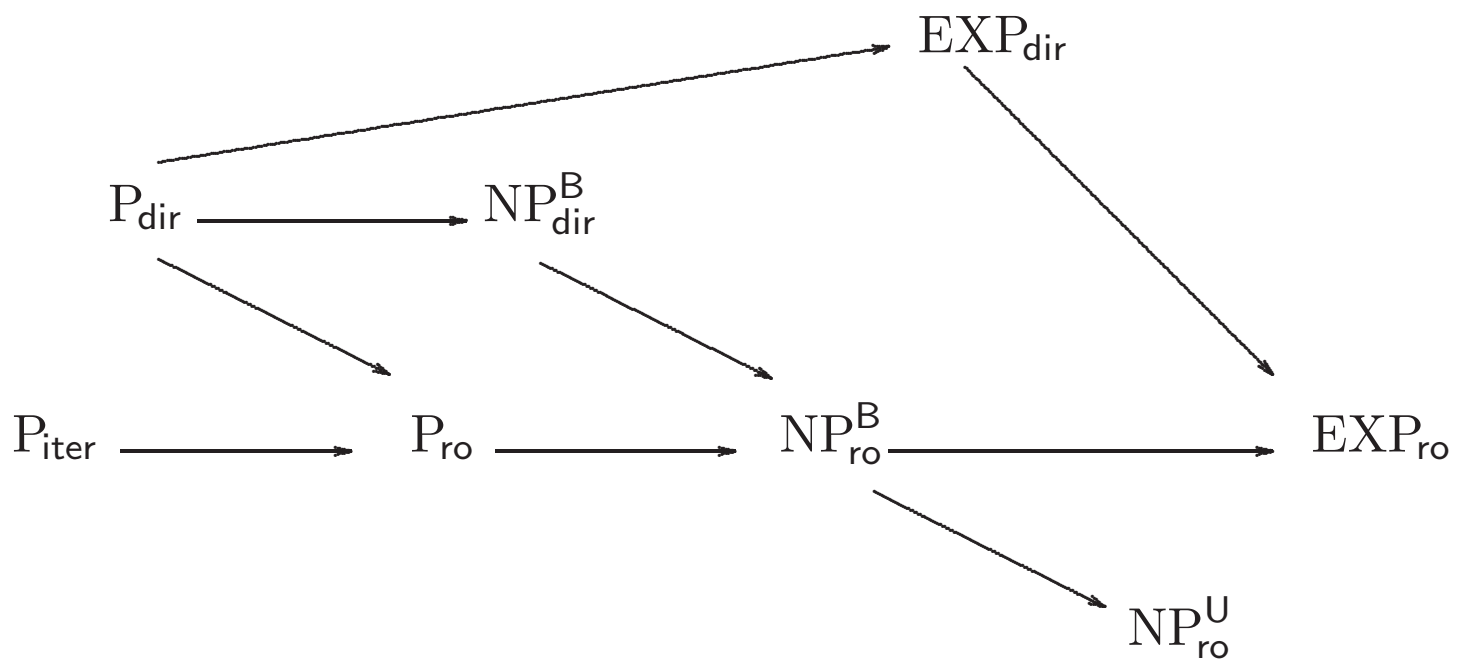

Decision problems, a notion of input size, finite-precision machines, and the cost of their computations, are all introduced in Section 3. The classes $\mathrm{P}_{\mathsf{ro}}, \mathrm{P}_{\mathsf{dir}}$ and $\mathrm{P}_{\mathsf{iter}}$ are described in Section 4. The classes $\mathrm{NP}^{\mathsf{U}}_{\mathsf{ro}}, \mathrm{NP}^{\mathsf{B}}_{\mathsf{ro}}, \mathrm{NP}^{\mathsf{U}}_{\mathsf{dir}}$ and $\mathrm{NP}^{\mathsf{B}}_{\mathsf{dir}}$ are examined in Section 5. Different versions of the *Circuit Feasibility Problem* are then shown to be complete in these classes, under appropriately defined reductions. The inclusion $\mathrm{NP}^{\mathsf{B}}_{\mathsf{ro}} \subset \mathrm{EXP}_{\mathsf{ro}}$ is shown in Section 6. Section 7 briefly discusses two issues related to our exposition: a relationship between condition and average complexity which is ubiquitous in the modern approach to condition, and a comparison with a Turing machine based approach to numerical computations.

Preceding these developments we spend some time overviewing the basic features of finite-precision computations and conditioning. For the latter, and in order to convey a certain ad-hoc character of condition numbers (cf. Remark 4 below), we are liberal with examples.

### 1.3 Previous and related work

To the best of our knowledge, this is the first exposition of a theory of NP-completeness that considers condition and finite-precision. Nonetheless, our development was only possible because substantial literature on related ideas was available.

To begin with, the condition-based accuracy analysis that goes back to the work of Turing [59] and von Neumann and Goldstine [60]. This analysis has since been pervasive in numerical linear algebra, and more recently began occurring in other subjects (linear optimization, polynomial computations, etc.).

Then there is the condition-based complexity analysis, which goes back to the analysis of the conjugate gradient method by Hestenes and Stiefel [34], and was more recently championed by Lenore Blum [4] and Steve Smale [55, 56]. The work of Jim Renegar on linear optimization [47, 48, 49] in the mid-90s is an outstanding example.

Thirdly, the theories of complexity that grew in the 1970s for discrete computations (see [30, 36, 43] for text expositions) and in the late 1980s for numerical problems (cf. [5]), which set the model for the kind of results one should look for in the finite-precision setting.

Finally, there are a few articles that, without attempting to build a theory, delve into the overlap of complexity and accuracy. Notably among them are an old result by Miller [42] establishing a trade-off between complexity and numerical stability for matrix multiplication, and a recent paper by Allender et al. [1] that relates a specific discrete problem with various aspects of numerical computations.

I cannot end this paragraph without mentioning another stream of research on the complexity of numerical computations. The point of departure now is the adoption of the Turing machine as computational model, together with the procedure of both inputting and outputting real numbers bit by bit. An excellent short account of this viewpoint is in [7]. Comprehensive expositions can be found in [38, 62]. We will briefly describe the differences with our approach in §7.2.

**Acknowledgments.** This paper stem from discussions with Gregorio Malajovich and Mike Shub. During the process of its writing I often communicated with both of them, and received constructive criticism for both definitions that did not capture, and proofs that did not establish, what they had to. I am greatly indebted to them.

# 2 Finite-precision, Condition and Stability

This section recounts the main features of finite-precision computations as they are performed in numerical analysis. The idea is not to propose a theory (this will be done in subsequent sections) but to acquaint readers possibly unfamiliar with these features and to motivate ensuing definitions.

## 2.1 Finite-precision computations

Numerical computations on a digital computer are supported by a representation of real numbers and their arithmetic. Because of the necessary finiteness of computer data real numbers are replaced by *approximations* and the ubiquitous form of these approximations are the so called *floating-point* numbers. We next briefly describe them, pointing to the reader that a comprehensive exposition of the subject is Chapter 2 in [35] (from where our short description has been extracted).

A *floating-point number system* $\mathbb{F} \subset \mathbb{R}$ is a set of real numbers $y$ having the form

$$y = \pm m \times \beta^{e-t}$$

where

**(1)** $\beta \in \mathbb{Z}$, $\beta \geq 2$, is the *base* of the system,

**(2)** $t \in \mathbb{Z}$, $t \geq 2$ is its *precision*,

**(3)** $e \in \mathbb{Z}$ satisfies $\mathsf{e}_{\min} \leq e \leq \mathsf{e}_{\max}$ (the *exponent range*) with $\mathsf{e}_{\min}, \mathsf{e}_{\max} \in \mathbb{Z}$, and $m \in \mathbb{Z}$ (the *mantissa*) satisfies $0 \le m \le \beta^t - 1$. We actually impose, to ensure a unique representation of $y$, that for $y \ne 0$ we have $\beta^{t-1} \le m \le \beta^t - 1$. This implies

$$y = \pm\beta^e \left(\frac{d_1}{\beta} + \frac{d_2}{\beta^2} + \cdots + \frac{d_t}{\beta^t}\right) \;=\; \pm\beta^e \times 0.d_1 d_2 \ldots d_t \tag{1}$$

with $0 \le d_i \le \beta - 1$ for all $i$ and $d_1 \ne 0$.

The non-zero elements $y$ of $\mathbb{F}$ satisfy

$$\beta^{\mathsf{e}_{\min}-1} \le |y| \le \beta^{\mathsf{e}_{\max}}(1 - \beta^{-t}). \tag{2}$$

The real numbers in these two intervals, along with $\{0\}$, make the *range* of $\mathbb{F}$, which we denote by $\mathsf{Range}(\mathbb{F})$. That is,

$$\mathsf{Range}(\mathbb{F}) := \big[-\beta^{\mathsf{e}_{\max}}(1-\beta^{-t}), -\beta^{\mathsf{e}_{\min}-1}\big] \;\cup\; \{0\} \;\cup\; \big[\beta^{\mathsf{e}_{\min}-1}, \beta^{\mathsf{e}_{\max}}(1-\beta^{-t})\big].$$

Associated to the system $\mathbb{F}$ there is a *rounding function* $\mathsf{fl} : \mathbb{R} \to \mathbb{F}$ which maps each real $x$ in the range of $\mathbb{F}$ to an element $\mathsf{fl}(x)$ in $\mathbb{F}$ closest to $x$ (there are several ways to break ties whose nature is of no consequence to our development; the interested reader can see the Notes and References of Chapter 2 in [35]). If $x$ is not in the range of $\mathbb{F}$ then it is either too large in absolute value ($|x| > \beta^{\mathsf{e}_{\max}}(1-\beta^{-t})$) or too small ($0 < |x| < \beta^{\mathsf{e}_{\min}-1}$). We talk about *overflow* or *underflow*, respectively. Different implementations of $\mathsf{fl}$ treat these cases in different manners, a common one letting $\mathsf{fl}(x)$ map $x$ to the nearest non-zero element in $\mathbb{F}$.

The *unit roundoff* of $\mathbb{F}$ is defined to be $\mathsf{u}_{\mathsf{mach}} := \frac{1}{2}\beta^{1-t}$. "It is the most useful quantity associated with $\mathbb{F}$ and is ubiquitous in the world of rounding error analysis" [35, p. 42]. It satisfies the following:

$$\text{for all } x \in \mathsf{Range}(\mathbb{F}),\ \mathsf{fl}(x) = x(1+\delta) \text{ for some } \delta \text{ with } |\delta| < \mathsf{u}_{\mathsf{mach}}. \tag{3}$$

Arithmetic in $\mathbb{F}$, in the *standard model*, is performed by first computing on $\mathbb{Q}$ the exact result of an operation and then applying $\mathsf{fl}$ to it. This defines, for any operation $\circ \in \{+, -, \times, /\}$ a corresponding operation

$$\tilde{\circ} : \mathbb{F} \times \mathbb{F} \to \mathbb{F}$$

which satisfies, for all $x, y \in \mathbb{F}$,

$$x \circ y \in \mathsf{Range}(\mathbb{F}) \Rightarrow x \tilde{\circ} y = (x \circ y)(1+\delta) \text{ for some } \delta \text{ with } |\delta| < \mathsf{u}_{\mathsf{mach}}. \tag{4}$$

**Remark 1** There is a qualitative difference between the bounds required for mantissas ($t < \infty$) and exponents ($|\mathsf{e}_{\min}|, \mathsf{e}_{\max} < \infty$) in a floating-point number system. For all real numbers $x > 0$ the exponent $e$ needed to represent $x$ in the interval $(\beta^{t-1}, \beta^t - 1)\beta^{e-t}$ is finite. In contrast with this, for almost all real numbers $x$ one must have $t = \infty$ if one wants that $\mathsf{fl}(x) = x$. This feature, together with the fact that over and underflow are rare when compared with the all-pervasive presence of rounding errors, is at the origin of the fact that, almost invariably, theoretical analyses of roundoff assume a floating-point system without bounds for the exponents. That is, a system where (3) and (4) hold true without requiring $x \in \mathsf{Range}(\mathbb{F})$ (or, equivalently, where $\mathsf{Range}(\mathbb{F}) = \mathbb{R}$). We will refer to such a system as having *unrestricted exponents.* Most accuracy analyses in the literature (e.g., all those cited in this paper) assume unrestricted exponents.

### 2.2 A helpful equality

The sequencing of arithmetic operations in the execution of an algorithm entails the accumulation of errors given by equation (4) and with it, the occurrence of products of quantities of the form $(1+\delta)$ with $|\delta| \leq \mathsf{u}_{\mathsf{mach}}$. The following result (see [35, Lemma 3.1]) deals with these products.

**Lemma 1** *If* $|\delta_i| \leq \mathsf{u}_{\mathsf{mach}}$ *and* $\rho_i = \pm 1$ *for* $i = 1, \ldots, n$*, and* $n\mathsf{u}_{\mathsf{mach}} < 1$ *then*

$$\prod_{i=1}^{n} (1+\delta_i)^{\rho_i} = 1 + \theta_n$$

*where* $\theta_n$ *is a real number satisfying*

$$|\theta_n| \leq \frac{nu}{1-nu} =: \gamma_n. \qquad \square$$

### 2.3 Stability and condition

The possible effect of roundoff errors on a computation raised the attention of the founding figures of modern numerical analysis. Both Turing in the U.K. and von Neumann and Goldstine in the U.S. considered this effect for the case of linear equation solving and attempted a quantitative explanation [59, 60]. The story of the former is nicely described by Wilkinson in his 1970's Turing Lecture [63]. What follows is a brief exposition of the ideas introduced by them and their subsequent extensions.

Errors in a finite-precision algorithm $\mathcal{A}$ computing a function $\varphi : U \subset \mathbb{R}^n \to \mathbb{R}^m$ will accumulate and $\mathcal{A}$ will return, on input $x \in \mathbb{R}^n$, a point $\varphi^{\mathcal{A}}(x)$ different from $\varphi(x)$. The extent of this difference can be measured by the *normwise relative error*

$$\mathsf{RelError}(\varphi^{\mathcal{A}}(x)) := \frac{\|\varphi^{\mathcal{A}}(x) - \varphi(x)\|}{\|\varphi(x)\|}. \tag{5}$$

An approach that has often worked to estimate this error passes through showing that

$$\varphi^{\mathcal{A}}(x) = \varphi(\widetilde{x}) \tag{6}$$

for a point $\widetilde{x} \in \mathbb{R}^n$ sufficiently close to $x$, say, satisfying that

$$\mathsf{RelError}(\widetilde{x}) := \frac{\|\widetilde{x} - x\|}{\|x\|} \leq \mathsf{u_{mach}}\, g(n, m) \tag{7}$$

where $g(n, m)$ grows slowly with $n$ and $m$ (e.g., as a low degree polynomial). If such a result —known as *backward-error analysis*— is possible, the relative error in (5) can be estimated from the knowledge of the *(relative, normwise) condition number* of $x$

$$\mathsf{cond}^{\varphi}(x) := \lim_{\delta \to 0} \sup_{\mathsf{RelError}(\widetilde{x}) \leq \delta} \frac{\mathsf{RelError}(\varphi(\widetilde{x}))}{\delta}. \tag{8}$$

Equation (8) shows the usual understanding of a condition number as "the worst possible magnification of the error, in the value of $\varphi(x)$, produced by a small perturbation of the data $x$." It follows from (5–8) that

$$\mathsf{RelError}(\varphi^{\mathcal{A}}(x)) \leq \mathsf{u_{mach}}\, g(n, m) \mathsf{cond}^{\varphi}(x) + o(\mathsf{u_{mach}}).$$

It is important to note here that a backward-error analysis is not always possible. In these cases, one needs to obtain bounds for $\mathsf{RelError}(\varphi^{\mathcal{A}}(x))$ with a more direct (and usually laborious) approach referred to as *forward-error analysis*.

**Example 1** For the case of matrix inversion we have the function $A \overset{\varphi}{\mapsto} A^{-1}$ from the set of invertible matrices to $\mathbb{R}^{n \times n}$. Under the assumption of unrestricted exponents (cf. Remark 1 above), the corresponding condition number $\mathsf{cond}^{\varphi}(A)$ is exactly $\|A\|\|A^{-1}\|$, a quantity usually denoted by $\kappa(A)$. Similar bounds hold for the problem of linear equation solving, $(A, b) \overset{\psi}{\mapsto} x = A^{-1}b$, just that now we only have the bounds $\kappa(A) \leq \mathsf{cond}^{\psi}(A, b) \leq 2\kappa(A)$. Also, in this case, a backward error analysis shows that the computed (using Householder QR decomposition) solution $\widetilde{x}$ satisfies, for some constant $C$,

$$\frac{\|\widetilde{x} - x\|}{\|x\|} \leq Cn^3 \mathsf{u_{mach}}\, \mathsf{cond}^{\varphi}(A) + o(\mathsf{u_{mach}}). \tag{9}$$

**Remark 2** Soon after the introduction of $\kappa(A)$ for error analysis, Hestenes and Stiefel [34] showed that this quantiy also played a role in complexity analyses. More precisely, they showed that the number of iterations of the conjugate gradient method (assuming infinite precision) needed to ensure that the current approximation to the solution of a linear system attained a given accuracy is proportional to $\sqrt{\kappa(A)}$.

A goal of this section is to show that, in spite of the rigor of the definition in (8) the idea of condition has a bit of an ad-hoc character (we will return to this idea in Remark 4 below). The next two examples show a first gradual departure of condition as given in (8).

**Example 2** Given a matrix $A \in \mathbb{R}^{m\times n}$ such that $K(A) := \{y \in \mathbb{R}^n \mid Ay \geq 0\} \neq \{0\}$ we want to find a point in $K(A) \setminus \{0\}$.

This problem does not fit the framework above in the sense that the function $\varphi$ is not well-defined: any point $y$ in $K(A)$ would do. In 1980 Goffin [29] analyzed the cost of a procedure to find one such $y$ in terms of the "best conditioned" point in $K(A)$. For $y \in \mathbb{R}^m$ one defines

$$\rho(A, y) := \min_{i\leq m} \frac{a_i \cdot y}{\|a_i\|\|y\|},$$

then

$$\rho(A) := \sup_{y\in K(A)} \rho(A, y)$$

and finally

$$\mathscr{C}(A) := \frac{1}{\rho(A)}.$$

Goffin's complexity analysis is in terms of $\mathscr{C}(A)$ (in addition to $n$ and $m$).

**Example 3** Let $d_1, \ldots, d_n$ be positive integers and $\mathbf{d} = (d_1, \ldots, d_n)$. We denote by $\mathcal{H}_{\mathbf{d}}$ the complex vector space of systems $f = (f_1, \ldots, f_n)$ with $f_i \in \mathbb{C}[X_0, \ldots, X_n]$ homogeneous of degree $d_i$. The problem is to compute (i.e., to approximate) a zero of $f \in \mathcal{H}_{\mathbf{d}}$. Generically, such a system $f$ has $\mathcal{D} := d_1 \cdot \ldots \cdot d_n$ different zeros in complex projective space $\mathbb{P}^n$. For each one of them, Shub and Smale have characterized the value of $\mathsf{cond}^\zeta(f)$ (here $\zeta$ is the selected zero) to be

$$\mu(f, \zeta) := \|f\| \left\| Df(\zeta)^{-1}_{|T_\zeta} \mathsf{diag}(\|\zeta\|^{d_i-1}) \right\|$$

where $T_\zeta$ is the tangent space at $\zeta$ and the inverse is of the restriction of $Df(\zeta)$ to this tangent space. Also, here $\mathsf{diag}(x_i)$ denotes the diagonal matrix with entries $\{x_i\}$, and $\|f\|$ is the norm induced by Weyl's Hermitian product on $\mathcal{H}_{\mathbf{d}}$. A variation of this quantity,

$$\mu_{\mathsf{norm}}(f, \zeta) := \|f\| \left\| Df(\zeta)^{-1}_{|T_\zeta} \mathsf{diag}(\sqrt{d_i}\|\zeta\|^{d_i-1}) \right\|$$

has the advantage of being unitarily invariant and has been used is several analyses of algorithms for approximating zeros [50, 51, 52, 54, 53, 2, 3, 9]. For this problem, and as in the previous example, since the data is the system $f$ and no zero $\zeta$ is specified (any one would do) a condition number $\mu(f)$ is defined in terms of the collection $\{\mu_{\mathsf{norm}}(f, \zeta)\}_{\zeta | f(\zeta)=0}$.

**Example 4** Let $a > 0$. We want to approximate $\sqrt{a}$ by using Hero's method (from Hero of Alexandria, although the method was already known to the Babylonians).

Assume, for the time being, that $a \in [1, 4)$ and take $x_0 = \frac{5}{2}$ so that $x_0 > \sqrt{a}$. Let $x_{k+1} = H_a(x_k)$ where

$$H_a(x) := \frac{1}{2}\Big(x + \frac{a}{x}\Big).$$

Because of roundoff errors we actually compute a sequence $\widetilde{x}_{k+1}$ and it is not difficult to show that, for some small constant $C$,

$$0 < \frac{\widetilde{x}_k - \sqrt{a}}{\sqrt{a}} \leq \frac{3}{2^{k+1}} + C\,\mathsf{u}_{\mathsf{mach}}.$$

It follows that to ensure that $\frac{\widetilde{x}_k - \sqrt{a}}{\sqrt{a}} < \varepsilon$ it is enough to have both

$$k \geq |\log_2 \varepsilon| + 2 \qquad \text{and} \qquad \mathsf{u}_{\mathsf{mach}} \leq \frac{\varepsilon}{2C}. \tag{10}$$

To compute $\sqrt{a}$ with $a \geq 1$ arbitrary one computes $b \in [1,4)$ and $q \in \mathbb{N}$ such that $a = b \cdot 4^q$, and then $\sqrt{a} = \sqrt{b}\,2^q$. The case $a \in (0,1)$ is dealt with using that $\sqrt{a^{-1}} = (\sqrt{a})^{-1}$.

When $a \in [1,4)$, the requirements (10) that ensure a relative error bounded by $\varepsilon$ are independent of $a$. This is consistent with the fact that, for all $x > 0$, the condition number $\mathsf{cond}^{\sqrt{\ }}(x)$ given by (8) is constant (and equal to $\frac{1}{2}$). For arbitrary $a > 0$, in contrast, the scaling process, i.e., the computation of $b$ and $q$ above, depends on the magnitude of $a$, both in terms of complexity (the value of $k$ grows linearly with $\log q$) and accuracy (the log of $\mathsf{u}_{\mathsf{mach}}^{-1}$ also grows linearly with $\log q$).

## 2.4 The Condition Number Theorem

A common feature of the problems in Examples 1 to 3 is the existence of a subset of data instances at which the condition number is not well-defined (or, more precisely, takes the value $\infty$). These data are considered *ill-posed* with respect to the problem at hand in the sense that no amount of (finite) precision will guarantee a given accuracy in the output. An old result [26] related to Example 1, nowadays known as *Condition Number Theorem* (CNT in what follows) shows that, if $\Sigma$ denotes the set of non-invertible matrices (which are the ill-posed matrices for the problem of matrix inversion or linear equation solving), then

$$\kappa(A) = \frac{\|A\|}{d(A,\Sigma)}.$$

Here $\|A\|$ denotes the spectral norm, and $d(A,\Sigma)$ refers to the distance induced by this norm. A systematic search for relations between condition and distance to ill-posedness was campaigned by Jim Demmel [24]. These relations are nowadays frequently established.

## 2.5 Finite-valued problems

For a finite-valued problem, that is, one given by a function $\varphi : \mathbb{R}^n \to F$ where $F$ is a finite set (say $F = \{y_1, \ldots, y_k\}$) the quantity $\mathsf{cond}^{\varphi}(x)$ is of little use. It is immediate to check that $\mathsf{cond}^{\varphi}(x) = \infty$ when $x$ is in the boundary of some $S_j := \{x \in \mathbb{R}^n \mid \varphi(x) = y_j\}$, for $j = 1, \ldots, k$, and that $\mathsf{cond}^{\varphi}(x) = 0$ otherwise.

The family of boundaries between the sets $S_j$ is composed of data $x$ for which there is no hope that a finite-precision computation with input $x$ will yield a reliable output. Elements in these boundaries are considered to be *ill-posed*.

An idea championed by Jim Renegar [47, 48, 49] is to define condition, for finite-valued problems, as the (relativized) inverse to the distance to ill-posedness (i.e., to impose a CNT). But other approaches to define condition for these problems have been used as well.

**Example 5** One can turn the problem in Example 2 into a feasibility problem: given $A \in \mathbb{R}^{m\times n}$ decide whether $K(A) \neq \{0\}$. This problem can be solved with finite precision and both the accuracy and the complexity of the algorithm are (nicely) bounded in terms of a simple extension of $\mathscr{C}(A)$ (for not necessarily feasible matrices $A$). One takes

$$\rho(A) := \sup_{\|y\|=1} \rho(A, y)$$

and

$$\mathscr{C}(A) := \frac{1}{|\rho(A)|}.$$

Note that $\rho(A) \geq 0$ if and only if $A$ is feasible and $A$ is ill-posed precisely when $\rho(A) = 0$. This extension was done in [12] where it was proved that $\mathscr{C}(A)$ satisfies a CNT, namely, that $\mathscr{C}(A) = \frac{\|A\|_{12}}{d_{12}(A,\Sigma)}$ (here $\| \;\|_{12}$ is the 1-2 operator norm, $d_{12}$ its associated distance, and $\Sigma$ the boundary between the sets of feasible and infeasible matrices $A$). This extension was then used to analyze interior-point methods for the feasibility problem (see [21] and Chapters 6, 7, 9, and 10 in [10]).

A variety of other condition measures have been proposed for this feasibility problem. A description of a few of them (with some comparisons) appears in [13].

**Example 6** The space $\mathcal{H}_{\mathbf{d}}^{\mathbb{R}}$ is as $\mathcal{H}_{\mathbf{d}}$ but with real coefficients. The problem now is, given $f \in \mathcal{H}_{\mathbf{d}}^{\mathbb{R}}$, count the number of real projective zeros of $f$. This is a finite-valued problem. A finite-precision algorithm that solves it is described in [19] and both its complexity and accuracy analyzed in terms of the condition number

$$\kappa(f) := \max_{x\in\mathbb{S}^n} \frac{\|f\|}{(\|f\|^2 \mu_{\mathsf{norm}}(f,x)^{-2} + \|f(x)\|_2^2)^{\frac{1}{2}}}.$$

Here $\|f\|$ is the norm induced by Weyl's inner product on $\mathcal{H}_{\mathbf{d}}^{\mathbb{R}}$. In [20] it is shown that $\kappa(f)$ satisfies a CNT. Indeed, a system $f$ is ill-posed when arbitrary small perturbations can change the number of its real zeros. Denote by $\Sigma$ the set of ill-posed systems. Then $\kappa(f) = \frac{\|f\|}{d(f,\Sigma)}$.

**Example 7** Let $\mathcal{H}^{\mathbb{R}}_{[\mathbf{d},m]}$ denote the space of systems of $m$ homogeneous polynomials in $n+1$ variables with real coefficients. We want to determine whether one such system $f$ is feasible. That is, whether there exists $x \in \mathbb{S}^n$ such that $f(x) = 0$. Here $\mathbb{S}^n$ denotes the unit sphere in $\mathbb{R}^{n+1}$. An algorithm for solving this problem was given in [22] and analyzed in terms of

$$\kappa_{\mathsf{feas}}(f) := \begin{cases} \min\limits_{\zeta \in Z_{\mathbb{S}}(f)} \mu_{\dagger}(f,\zeta) & \text{if } Z_{\mathbb{S}}(f) \neq \emptyset \\ \max\limits_{\zeta \in \mathbb{S}^n} \dfrac{\|f\|}{\|f(x)\|} & \text{otherwise.} \end{cases}$$

Here $Z_{\mathbb{S}}(f)$ denotes the zero set of $f$ on $\mathbb{S}^n$ and $\mu_{\dagger}(f,\zeta)$ a version of $\mu_{\mathsf{norm}}(f,\zeta)$ for overdetermined systems defined with the help of the Moore-Penrose inverse.

**Remark 3** For the condition numbers $\mathscr{C}(A)$ and $\kappa(f)$ in Examples 5 and 6 we have that the condition is $\infty$ if and only if the data is in the boundary between sets of the form $S_j = \{x \in \mathbb{R}^n \mid \varphi(x) = y_j\}$ (inputs with a specific output).

This is not the case for $\kappa_{\mathsf{feas}}(f)$ in Example 7. This condition number takes the value infinity at the ill-posed systems $f$ but it does so at other systems as well (for instance on the case $m = 1$, $n = 1$, $d = 3$ and the polynomial $X_1^3$).

We still want to say that inputs $f$ for which $\kappa_{\mathsf{feas}}(f) = \infty$ are ill-posed. To distinguish between them and those in the boundaries between sets $S_j$ we will, using an expression due to Jim Renegar [46], call the latter *definitely ill-posed*.

**Remark 4** We close this section returning to the nature of condition numbers. Already the definition in (8) depends on a number of choices. For instance, the selection of a particular norm, or the way of measuring errors, which needs not be normwise (as in (5) and (7)) but may be componentwise instead. On top of this, we just saw that for finite-valued problems the quantity defined in (8) is of no practical use and other forms of condition need to be defined. Examples 5, 6, and 7 show that there is not a single, all-purpose, choice here either.

On top of all this, the result of Hestenes and Stiefel mentioned in Remark 2 triggered the emergence of an assortment of measures associated to data for various problems which were used for the complexity analysis of iterative algorithms. Almost invariably these measures were also referred to as condition numbers.

# 3 Decision Problems and Finite-precision Machines

We formally define in this section both the class of problems we will deal with and the model for the machines solving them.

## 3.1 Decision problems

Among the finite-valued problems the class of decision problems deserves emphasis. These are problems with only two outputs and are usually stated as a question on the input data which may have as answer either `Yes` or `No`. In the rest of this paper we will focus on decision problems.

Before giving a formal definition of decision problems we note that natural objects occurring in the theory (circuits, polynomials, machines, ...) have discrete and continuous components. Consistently with this division, algorithms perform computations both with real and discrete data (and computer languages such as `C`, `Fortran`, or `Mathlab`, distinguish between floating-point and integer numbers and implement arithmetics for both). To reflect this situation we will consider data in the product

$$\mathcal{I} := \{0,1\}^\infty \times \mathbb{R}^\infty$$

and define decision problems to appropriately take into account this structure. Here

$$\mathbb{R}^\infty := \bigsqcup_{i=0}^{\infty} \mathbb{R}^i$$

where the union is disjoint and $\mathbb{R}^0$ is an empty symbol so that $\{0,1\}^\infty \times \mathbb{R}^0 \simeq \{0,1\}^\infty$. Similarly for $\{0,1\}^\infty$.

**Definition 1** A *decision problem* is a pair $(A, \boldsymbol{\mu})$ where $A \subset \mathcal{I}$ and $\boldsymbol{\mu} : \mathcal{I} \to [1, \infty]$. Here $\boldsymbol{\mu}$ is the *condition number.*

We denote by $\Sigma$ the set $\{(u, x) \in \mathcal{I} \mid \boldsymbol{\mu}(u, x) = \infty\}$ and we say that elements in $\Sigma$ are *ill-posed.*

Examples 5 and 7 provide instances of decision problems.

**Remark 5** Different condition numbers for the same subset $A \subset \mathcal{I}$ define different decision problems. This is akin to the situation in classical (i.e., both discrete and infinite-precision BSS) complexity theory where different encodings of the intended input data define (sometimes radically) different problems.

For instance, to specify a univariate real polynomial in *dense encoding* we provide both its coefficients (the continuous data) and its degree (the discrete data). That is, we describe a polynomial $f = a_0 + a_1 X + \ldots + a_d X^d$ by using the array $[d, a_0, \ldots, a_d]$. Instead, if the encoding is *sparse*, we describe $f$ by the list $\{(i, a_i) \mid a_i \neq 0\}$. The size of the dense encoding may be exponentially larger than the size of the sparse one and consequently, the complexity of a problem (e.g., decide whether $f$ has a real root) may dramatically depend on which of these two encodings the input is given.

We will return to the issue of the arbitrariness of condition numbers in §7.1.

## 3.2 Finite-precision machines, input size, and computational cost

Briefly, a finite-precision machine is a BSS machine whose arithmetic is not exact but obeys the laws described in §2.1. We will not give a completely formal definition of them to avoid repeating a definition that nowadays is well-known (readers wishing to read such definition will find it in [6] or [5, §3.2]).

**Definition 2** A *finite-precision BSS machine* is a BSS machine performing finite-precision computations. To precisely define the latter, we fix a number $\mathsf{u}_{\mathsf{mach}} \in (0,1)$ (the *unit roundoff*) and let

$$\mathsf{k}_{\mathsf{mach}} := \left\lceil \log_2 \frac{1}{\mathsf{u}_{\mathsf{mach}}} \right\rceil .$$

In a $\mathsf{u}_{\mathsf{mach}}$-*computation*, built-in constants, input values, and the result of arithmetic operations, call any such number $z$, are systematically replaced by another real number $\mathsf{fl}(z)$ satisfying

$$\mathsf{fl}(z) = z(1+\delta) \quad \text{for some } |\delta| < \mathsf{u}_{\mathsf{mach}}. \tag{11}$$

For all $(u,x) \in \mathcal{I}$, we denote by $\mathsf{Comp}(M, \mathsf{u}_{\mathsf{mach}}, u, x)$ the set of all possible $\mathsf{u}_{\mathsf{mach}}$-computations of $M$ with input $(u,x)$.

We will refer to $\mathsf{k}_{\mathsf{mach}} \in \mathbb{N}$ as the *precision* of $M$.

Finally, we will assume that the machine $M$ has some special-purpose, read-only, registers storing the values of $\mathsf{k}_{\mathsf{mach}}$ (as an element in $\{0,1\}^\infty$) and $\mathsf{u}_{\mathsf{mach}}$.

**Remark 6** The definition above does not discriminate between real and discrete data. This is not necessary (and we have therefore proceeded with simplicity as a goal). Indeed, discrete data can be encoded by sequences of real numbers in $[0,+\infty)$. A number $x$ encodes 1 if $x > 0$ and encodes 0 otherwise. Furthermore, Turing machine computations can be simulated by finite-precision computations and these simulations are both robust —they are always correct, independently of the precision at hand— and efficient —the cost of the simulation is linear in the cost of the Turing's machine computation. We will assume both these encoding and simulations when talking about finite-precision machines taking inputs in $\{0,1\}^\infty \times \mathbb{R}^\infty$, refer to the data thus encoded as *discrete data*, and indicate as *discrete computations* those (simulations) performed on discrete data. It is crucial to keep in mind, and we repeat it here, that these computations are error-free.

**Remark 7** **(i)** Some of the details in the definition above are meant to allow for generality whereas some others are chosen to simplify the exposition. The notion of $\mathsf{u}_{\mathsf{mach}}$-computation mimics the finite-precision computations described in §2.1 under the assumtion of unrestricted exponents. But we do not impose a finite set $\mathbb{F}$ and actually allow for all possible outputs of $\mathsf{fl}$ in $\mathbb{R}$ as long as (11) is satisfied.

**(ii)** The inclusion of divisions as a basic arithmetic operation in BSS machines makes possible the occurrence of divisions by zero. We will assume (as done in [6]) that all divisions are preceeded by a test eliminating this possibility.

**(iii)** It is useful to think of the "machine program" and the constant $\mathsf{u}_{\mathsf{mach}}$ as separate entities. This allows one to consider the computations of a machine $M$ for different values of $\mathsf{u_{mach}}$. Recall, however, that the values of $\mathsf{k_{mach}}$ and $\mathsf{u_{mach}}$ are available to the program and that the machine may use these values during the computation.

**(iv)** A (somehow peculiar) consequence of the availability of $\mathsf{u_{mach}}$ to the machine program is the fact that some $\mathsf{u_{mach}}$-computations may not be $\mathsf{u'_{mach}}$-computations for some $\mathsf{u'_{mach}} > \mathsf{u_{mach}}$. The obvious example is a computation that tests an inequality of the form $z < \mathsf{u_{mach}}$.

**Definition 3** We will say that a computation is *exact* when all its arithmetic operations are performed error-free, that is, $\mathsf{fl}(z) = z$ in Definition 2.

Obviously, for every $\varepsilon > 0$ an exact computation is a possible $\varepsilon$-computation.

**Remark 8** It will occassionally be useful to talk about *infinite precision.* This amounts to set $\mathsf{k_{mach}} = \infty$ and $\mathsf{u_{mach}} = 0$. In this case, it is easy to see, we recover the standard BSS machines.

To deal with complexity we need to fix two yardsticks. A measure of size (for the inputs of a problem) and a measure of cost (for the computations solving this problem). Complexity is the dependence of the latter on the former.

We define the *length* of $(u, x) \in \{0,1\}^s \times \mathbb{R}^n \subset \mathcal{I}$, which we write as $\mathsf{length}(u, x)$, to be $s + n$. We define the *size* of $(u, x)$ as

$$\mathsf{size}(u, x) := \mathsf{length}(u, x) + \lceil \log_2 \boldsymbol{\mu}(u, x) \rceil.$$

Note that if $(u, x)$ is ill-posed then $\mathsf{size}(u, x) = \infty$ and that otherwise $\mathsf{size}(u, x) \in \mathbb{N}$. Also, that for pairs $(u, x)$ with $\boldsymbol{\mu}(u, x) = 1$ (in particular, for elements $u \in \{0,1\}^\infty$) we have $\mathsf{size}(u, x) = \mathsf{length}(u, x)$.

The computation of a BSS machine has a cost associated to it which is the number of steps performed before halting. We call this the *arithmetic cost* and, for the computation of a machine $M$ on input $(u, x) \in \mathcal{I}$, we denote it by $\mathsf{ar_cost}_M(u, x)$.

In addition to the arithmetic cost, we define the *accuracy cost* of a computation (of a machine $M$ on input $(u, x)$) to be the smallest value of $\mathsf{k_{mach}}$ guaranteeing a correct answer. This is commonly associated with the the cost in practice (measured in number of bit operations) of operating with floating-point numbers since, assuming the exponents of such numbers are moderately bounded, this cost is at most quadratic on $\mathsf{k_{mach}}$ for all the common implementations of floating-point arithmetic.

We can now deal with complexity.

### 3.3 Clocked computations

Complexity classes are usually defined by putting restrictions on computation resources (notably, running time) as a function of input length. Our situation demands for a more involved approach due to a number of features proper to it: size depends on condition as well as on length (and condition is not known a priori), output correctness depends on the machine precision, and total cost must depend on this machine precision as well (since the cost of arithmetic operations in practice does so). Definition 4 below intends to capture these features. It uses the common notion of time constructibility which we next recall.

A function $T : \mathbb{N} \to \mathbb{N}$ is *time constructible* when there exists a Turing machine that with input $n$ returns $T(n)$ in time $\mathcal{O}(T(n))$. Most of the functions used in complexity theory (e.g., polynomial and exponential functions) are time constructible.

**Definition 4** Let $\mathsf{Arith} : \mathbb{N} \times \mathbb{N} \to \mathbb{N}$ and $\mathsf{Prec} : \mathbb{N} \to \mathbb{N}$ be time constructible functions. We say that a decision problem $(S, \boldsymbol{\mu})$ is *solved with cost* $(\mathsf{Arith}, \mathsf{Prec})$ when there exists a finite-precision BSS machine $M$ satisfying the following. For every $(u, x) \in \mathcal{I}$ with $\boldsymbol{\mu}(u, x) < \infty$ the computation of $M$ with input $(u, x)$ satisfies

$$\mathsf{ar_cost}_M(u, x) \leq \mathsf{Arith}(\mathsf{length}(u, x), \mathsf{k_{mach}}),$$

and, if

$$\mathsf{k_{mach}} \geq \mathsf{Prec}(\mathsf{size}(u, x))$$

then all computations of $M$ correctly decide whether $(u, x) \in S$.

We observe that the machine $M$ in Definition 4 above needs not to halt for ill-posed inputs. In addition, we highlight two important features:

**(i)** Computations are clocked, i.e., their arithmetic cost is bounded by a function on two parameters immediately available: length of the input data and machine precision.

**(ii)** Computations are unreliable in the sense that there is no guarantee that the precision used is enough to ensure a correct output. Actually, correctness is not guaranteed even for exact computations.

Our basic deterministic complexity classes, $\mathrm{P}_{\mathsf{ro}}$ and $\mathrm{EXP}_{\mathsf{ro}}$ will be obtained by appropriately bounding $\mathsf{Arith}$ and $\mathsf{Prec}$ in Definition 4.

The evaluation of many common functions (e.g., a determinant) is done with clocked computations whose arithmetic cost, in general, depend only on the length of the input. The following example shows a general situation where, in contrast, this cost depends on $\mathsf{k_{mach}}$ as well.

**Example 8** We want to decide whether a continuous function (e.g., a polynomial) $f : [a, b] \to \mathbb{R}$ has a zero in the interval $[a, b]$. We consider as ill-posed any pair $(f, [a, b])$ on which all the zeros of $f$ are either at the endpoints of $[a, b]$ or are extrema of $f$. That is, such a pair is ill-posed if $f$ has a zero but does not change sign on $[a, b]$.

A simple scheme to decide this problem is to evaluate $f$ in a set $X = \{a = x_0, x_1, \ldots, x_n = b\}$, say of equally spaced points, and to reject if all the obtained values have the same sign. Common sense suggests that it is useless to have too many points on $X$ when $\mathsf{k_{mach}}$ is small. It also suggest that it is useless to have a large $\mathsf{k_{mach}}$ if the set $X$ has few points. The values of $\mathsf{k_{mach}}$ and $n$ will have to grow, as it were, in tandem (and here the fact that a machine can read its current $\mathsf{k_{mach}}$ is of the essence). When both values are low, we do not expect our scheme to correctly decide the existence of a zero of $f$. How large they need to be for the scheme to do so? This of course depends on the characteristics of $f$, and should be measured by a condition number. We will see a detailed version of this scheme in Theorem 4.

## 3.4 A hierarchy theorem

Early on the development of complexity theory it was proved that given more resources a Turing machine could solve more problems. These results were referred to as *hierarchy theorems* and the two best known are for time [32] and space [31].

In this paragraph we show a hierarchy theorem for precision.

**Proposition 1 (Precision Hierarchy Theorem)** *Let $T:\mathbb{N}\to\mathbb{N}$ be time constructible and $P_1, P_2 : \mathbb{R}_+ \to \mathbb{R}_+$ such that $P_2$ is continuous and increasing and $P_1 < \frac{P_2}{2}$. There exists a decision problem $(B, \boldsymbol{\mu})$ which can be decided with $\mathsf{ar_cost}(u,x) \le \mathcal{O}(T(\mathsf{length}(u,x)))$ and $\mathsf{k_{mach}} = P_2(\mathsf{size}(u,x)) + 3$, without reading the values of $\mathsf{k_{mach}}$ or $\mathsf{u_{mach}}$, but cannot be decided with $\mathsf{k_{mach}} = P_1(\mathsf{size}(u,x))$ (no matter the arithmetic cost).*

PROOF. Let $(B, \boldsymbol{\mu})$ given by the set

$$B := \left\{(n,x) \in \mathbb{N}\times\mathbb{R} \mid x \ge 0 \quad \text{and} \quad x^{2^{T(\mathsf{length}(n,x))}} \ge \frac{1}{2}\right\},$$

and the condition number

$$\boldsymbol{\mu}(n,x) := 2^{P_2^{-1}\left(\log\left(\frac{1}{\xi(n,x)}\right)\right)}$$

where

$$\xi(n,x) := \inf\{1, \varepsilon > 0 \mid \exists\delta, |\delta| \le \varepsilon \text{ s.t. } (n, x(1+\delta)) \in B \iff (n,x) \notin B\}.$$

The fact that $P_2$ is continuous and increasing allows us to use its inverse.

The definition of $\xi$ implies that, for all $(n,x)$, if $\mathsf{u_{mach}} > \xi(n,x)$ then for any possible machine deciding $B$ there are $\mathsf{u_{mach}}$-computations yielding a wrong answer for input $(n,x)$. Indeed, assume $(n,x) \in B$ and let $\delta$ be such that $|\delta| < \mathsf{u_{mach}}$ and $x(1+\delta) \notin B$. Then the computation that first rounds $x$ to $x(1+\delta)$ and then proceeds error-free, is an $\mathsf{u_{mach}}$-computation and returns that $(n,x) \notin B$ since this is the case for $(n, x(1+\delta))$. Likewise for the case $(n,x) \notin B$. It follows that the precision needed by any machine deciding $B$ satisfies $\mathsf{u_{mach}} \le \xi(n,x)$ for all input $(n,x)$. That is, we must have

$$\mathsf{k_{mach}} = \left\lceil \log \frac{1}{\mathsf{u_{mach}}} \right\rceil \ge \log \frac{1}{\xi(n,x)} = P_2(\log \boldsymbol{\mu}(n,x)).$$

Now consider any pair $(n,x)$ with $\mathsf{length}(n,x) \le \log \boldsymbol{\mu}(n,x)$. Then

$$\mathsf{size}(n,x) \le 2 \log \boldsymbol{\mu}(n,x)$$

and therefore, to decide such a pair we must have

$$\mathsf{k_{mach}} \ge P_2(\log \boldsymbol{\mu}(n,x)) \ge P_2\left(\frac{\mathsf{size}(n,x)}{2}\right) > P_1(\mathsf{size}(n,x)).$$

This shows that (independently of arithmetic cost considerations) $(B, \boldsymbol{\mu})$ cannot be decided with $\mathsf{k_{mach}} \le P_1(\mathsf{size}(n,x))$.

To conclude, we will show that $(B, \boldsymbol{\mu})$ can be solved with the claimed cost bounds. To do so, we consider the algorithm that computes $x^{2^{T(\mathsf{length}(n,x))}}$ by repeated squaring and use a simple backward error argument.

The algorithm first computes $t := T(\mathsf{length}(n,x))$ (note, this is a discrete computation whose cost is $\mathcal{O}(T(\mathsf{length}(n,x))$ since $T$ is time constructible) and then performs $t$ multiplications. Its arithmetic cost is therefore $\mathcal{O}(T(\mathsf{length}(n,x))$. In addition, using Lemma 1, it is easy to see that the computed value $q$ is of the form

$$x^{2^t}(1 + \theta_{2^{t+1}-1})$$

and therefore, of the form

$$x^{2^t}(1+\theta_2)^{2^t} = (x(1+\theta_2))^{2^t}$$

where, we recall, $\theta_2 \in \mathbb{R}$ satisfies $|\theta_2| \le \gamma_2 = \frac{2\mathsf{u_{mach}}}{1-2\mathsf{u_{mach}}} \le 3\mathsf{u_{mach}}$ if $\mathsf{u_{mach}} \le \frac{1}{6}$.

Take $\mathsf{k_{mach}} := P_2(\mathsf{size}(n,x)) + 3$. Then,

$$\mathsf{u_{mach}} = 2^{-P_2(\mathsf{size}(n,x))+3} \le \frac{1}{8}\, 2^{-P_2(\log \boldsymbol{\mu}(n,x))} = \frac{\xi(n,x)}{8}.$$

Our choice of $\mathsf{k_{mach}}$ also implies $\mathsf{u_{mach}} \le \frac{1}{6}$ and therefore, that $|\theta_2| \le 3\mathsf{u_{mach}} < \xi(n,x)$. It follows from the definition of $\xi(n,x)$ that $(x(1+\theta_2))^{2^t} \ge \frac{1}{2}$ if and only if $x^{2^t} \ge \frac{1}{2}$ and therefore, that the machine correctly decides the pair $(n,x)$. □

# 4 Polynomial Cost

We focus in this section in polynomial cost. We first define the general class capturing this notion and then proceed to describe the subclasses $\mathrm{P_{dir}}$ and $\mathrm{P_{iter}}$.

### 4.1 General polynomial time: the class $\mathrm{P}_{\mathsf{ro}}$

**Definition 5** A decision problem $(S, \boldsymbol{\mu})$ belongs to $\mathrm{P}_{\mathsf{ro}}$ (*roundoff polynomial cost*) when there exists a finite-precision BSS machine $M$ solving $S$ with cost $(\mathsf{Arith}, \mathsf{Prec})$ and such that

**(i)** $\mathsf{Prec}$ is bounded by a polynomial function, and

**(ii)** the function $\mathsf{Arith}(\mathsf{length}(u,x), \mathsf{Prec}(\mathsf{size}(u,x)))$ is bounded by a polynomial in $\mathsf{size}(u,x)$, for all $(u,x) \in \mathcal{I}$.

We note that the polynomial bound on $\mathsf{Prec}$ is satisfied whenever a bound of the form

$$\mathsf{u}_{\mathsf{mach}} \leq \frac{E}{2^{(\mathsf{length}(u,x))^r} \boldsymbol{\mu}(u,x)^s},$$

for some constants $E, r, s > 0$, ensures that $M$ correctly decides whether $(u,x) \in S$. It is this kind of expression which is commonly found in the literature.

**Remark 9** **(i)** Definition 5 is somehow tortuous, and a few remarks may help to understand the issues at hand.

The fact that $\mathsf{Prec}$ is bounded by a polynomial guarantees that the precision required by the algorithm is at most polynomial in the input size and hence, that the (Turing) cost of each arithmetic operation is so.

Furthermore, the bound in (ii) implies that the arithmetic cost with this required precision is also polynomially bounded in the size of the input.

But the definition allows for various ways to achieve this combination. In the simplest, both $\mathsf{Prec}$ and $\mathsf{Arith}$ are polynomials in their arguments. A different possibility would allow a smaller bound for $\mathsf{Prec}$, say logarithmic, and in this case the dependance of $\mathsf{Arith}$ on its second variable could be exponential and still have (i) holding true. In this case we say that $(S, \boldsymbol{\mu})$ can be solved with *logarithmic precision*.

**(ii)** It is clear that the quality of being in $\mathrm{P}_{\mathsf{ro}}$, for the problem of deciding a subset $S \subset \mathcal{I}$, depends on the condition number $\boldsymbol{\mu}$ associated to $S$. As we mentioned in Remark 5, this is akin to the situation in classical complexity theory where different encodings of the intended input may affect the membership of the problem to the class $\mathsf{P}$ (over $\{0,1\}$) or $\mathrm{P}_{\mathbb{R}}$ (over $\mathbb{R}$).

The choice of a particular condition number (just as the choice of a particular encoding) is outside the theory. In the case of condition numbers, a probabilistic analysis allows one to compare complexity bounds (for possibly different algorithms, in terms of possibly different condition numbers) by taking expectations and expressing expected values of functions of condition numbers in terms of the length of the input. We will return to this theme in §7.1.

## 4.2 Fixed- and variable-precision

A cursory look at various textbooks in numerical analysis shows the existence of two categories of algorithms, referred to as *direct* and *iterative* (the table of contents of [25] is a case at hand). For instance, for linear equation solving, Gauss elimination is a direct method whereas Jacobi's method is iterative, and for linear programming, the same disctinction applies to the Simplex and Interior Point methods, respectively.

This grouping is strongly related to another one, based on whether the precision $\mathsf{u_{mach}}$ an algorithm works with needs to be considered as fixed or can be increased during the execution of the algorithm. In the first case one can expect correct outputs only for sufficiently well-conditioned inputs, whereas the goal in the second is to ensure correct outputs for all well-posed inputs, at the price of an undeterminate halting time.

For the case of functional (as opposed to decisional) problems, the difference can be discerned on examples we have already seen. In Example 1 we solve a system $Ax = b$ with Householder QR decomposition, which requires $\mathcal{O}(n^3)$ operations, a bound depending on the dimension of $A$ only. The computed solution $\widetilde{x}$ satisfies the relative error bound (9), whose right-hand side cannot be estimated without knowledge of the condition number $\kappa(A)$. Hence, we are uncertain about the magnitude of this error.

In Example 4, instead, given $\varepsilon > 0$, we can guarantee a relative error at most $\varepsilon$ for the computed approximation $\widetilde{x_k}$ of $\sqrt{a}$ provided $k \geq |\log_2 \varepsilon| + 2$ and $\mathsf{u_{mach}} \leq D\varepsilon$ for a constant $D$. Now both the arithmetic cost and the reequired precision will increase with a decrease on $\varepsilon$ but, if we are allowed to adjust $\mathsf{u_{mach}}$, we can ensure that the relative error is bounded by this $\varepsilon$.

The distinction between fixed and variable precision, even though less common in the literature, will serve us to define two natural subclasses of $\mathrm{P}_{\mathsf{ro}}$.

## 4.3 Fixed-precision: the class $\mathrm{P}_{\mathsf{dir}}$

**Definition 6** A decision problem $(S, \boldsymbol{\mu})$ belongs to $\mathrm{P}_{\mathsf{dir}}$ (*direct polynomial cost*) when there exists a finite-precision BSS machine $M$ satisfying the following. For every $(u, x) \in \mathcal{I}$ the computation of $M$ with input $(u, x)$ never reads the values of $\mathsf{k_{mach}}$ or $\mathsf{u_{mach}}$ and satisfies that

$$\mathsf{ar_cost}_M(u, x) \leq (\mathsf{length}(u, x))^{\mathcal{O}(1)},$$

and that, if

$$\mathsf{k_{mach}} \geq (\mathsf{size}(u, x))^{\mathcal{O}(1)}$$

then all computations of $M$ correctly decide whether $(u, x) \in S$.

If correctness is ensured as soon as $\mathsf{k_{mach}} \geq (\log \mathsf{size}(u, x))^{\mathcal{O}(1)}$ we say that $(S, \boldsymbol{\mu})$ can be solved with *logarithmic precision*.

**Remark 10** **(i)** Computations of machines in $\mathrm{P}_{\mathsf{dir}}$ always halt within a bounded number of steps since their time bound depend only on the input length. The output of such computation, however, may be incorrect and this will happen when the machine precision is insufficient.

**(ii)** If both $\mathsf{k}_{\mathsf{mach}} = \infty$ and $\mathsf{size}(u, x) = \infty$ we condider the second bound in Definition 6 to hold. This means that we can decide ill-posed inputs as long as we compute with infinite precision.

The following result is trivial.

**Proposition 2** *We have* $\mathrm{P}_{\mathsf{dir}} \subset \mathrm{P}_{\mathsf{ro}}$. □

The fact that a $\mathrm{P}_{\mathsf{dir}}$ machine never reads the values of $\mathsf{k}_{\mathsf{mach}}$ or $\mathsf{u}_{\mathsf{mach}}$ immediately yields the following result (compare with Remark 7(iv)).

**Proposition 3** *Let $M$ be a $\mathrm{P}_{\mathsf{dir}}$ machine and $\varepsilon < \delta$. Then every $\varepsilon$-computation of $M$ is a $\delta$-computation as well.* □

The notion of Boolean circuit plays a fundamental role in discrete complexity theory (see, e.g., [43, §4.3]). The same is true for algebraic circuits in algebraic complexity [5, §18.4], whose definition we recall next.

**Definition 7** An *algebraic circuit* is a connected, directed acyclic graph whose nodes have in-degree either 0, 2, or 3. Nodes with in-degree 0 are labeled either with a variable (we call them *input nodes*), or with a real constant (*constant nodes*). Nodes with in-degree 2 (called *arithmetic*) are labeled with an arithmetic operation in $\{+, -, \times, /\}$ . Nodes with in-degree 3 are called *selection nodes.* Nodes with out-degree 0 are called *output nodes.*

The following drawing gives an example of an algebraic circuit.

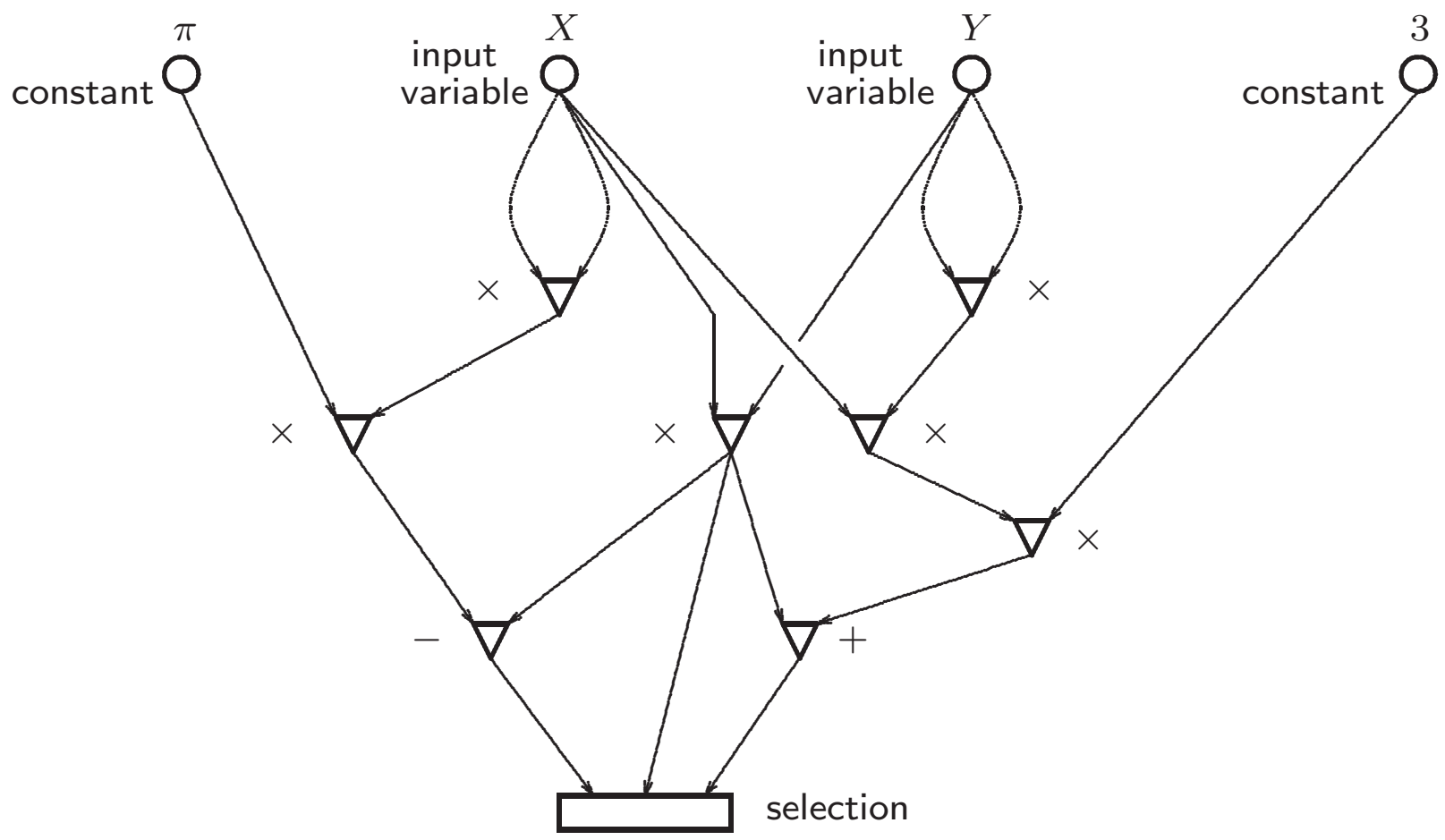

**Remark 11** As in Remark 7(ii), we will assume that all division nodes are preceeded by a test making sure that denominators are non-zero.

An algebraic circuit $\mathcal{C}$ with input variables $X_1, \ldots, X_n$ and $m$ output nodes has naturally associated to it the computation of a function $f_{\mathcal{C}} : \mathbb{R}^n \to \mathbb{R}^m$. Given a point $x \in \mathbb{R}^n$ this computation proceeds from input and constant nodes to output nodes by performing the arithmetic operations and the selections. For the latter, if the selection nodes has three parents $\xi, y, z$, the selection returns $y$ if $\xi < 0$ and $z$ if $\xi \geq 0$. We call this computation the *canonical procedure*.

The following diagram shows how the canonical evaluation is performed on the circuit drawn above for an input $(x, y) \in \mathbb{R}^2$.

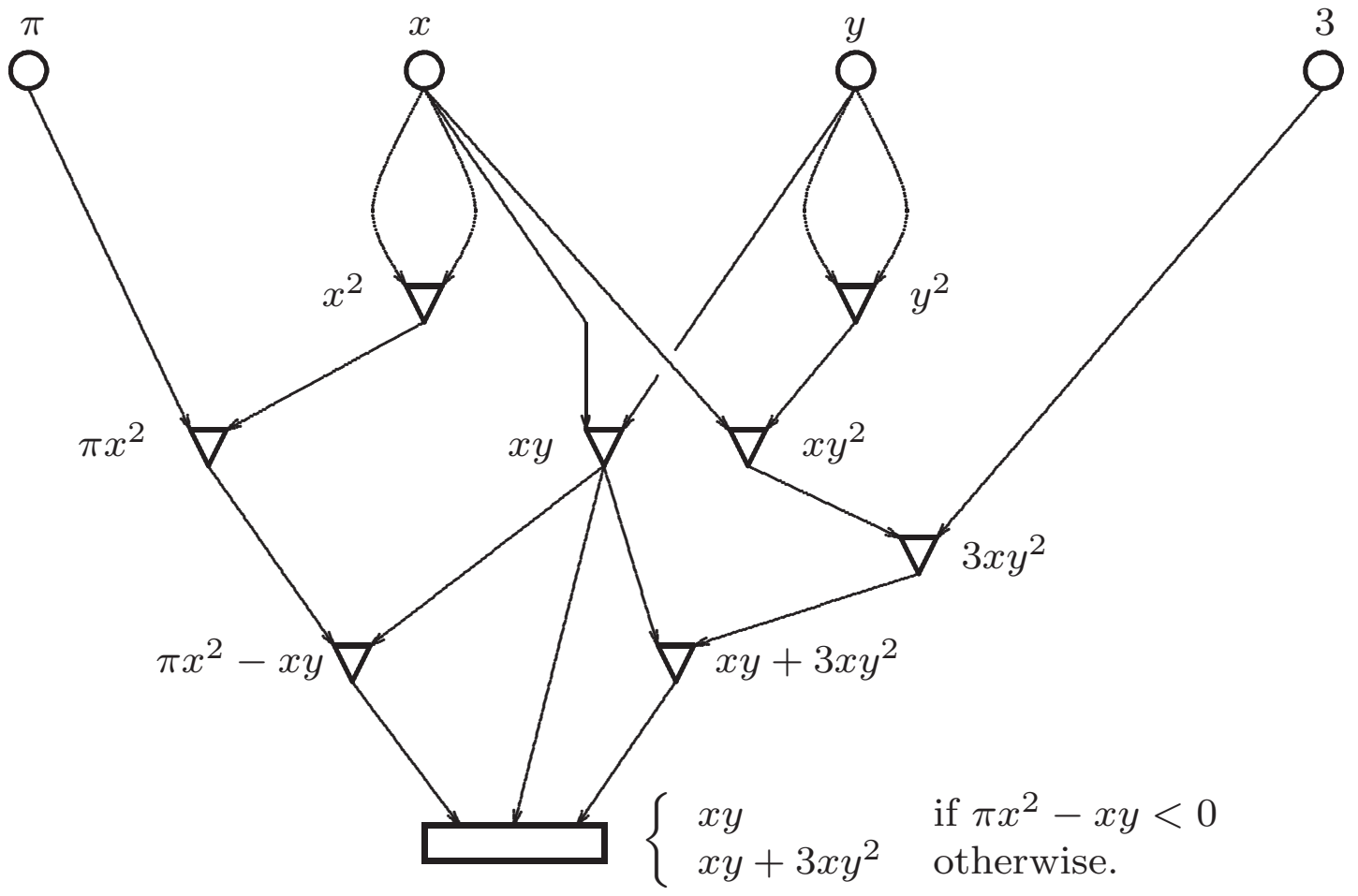

For $\varepsilon \in (0, 1)$ we define an $\varepsilon$-*evaluation* of $\mathcal{C}$ on input $x$ to be any finite-precision computation (as described in Definition 2) of the canonical procedure with $\mathsf{u_{mach}} = \varepsilon$. That is, any perturbation of this procedure in which the values of input variables, constants, or arithmetic nodes are multiplied by quantities of the form $(1 + \delta)$ with $|\delta| < \varepsilon$ (selections are performed error-free). The notion of an *exact* evaluation is similarly defined.

The following result is straightforward.

**Proposition 4** *If $\varepsilon < \delta$ then every $\varepsilon$-evaluation of $\mathcal{C}$ at $x \in \mathbb{R}^n$ is a $\delta$-evaluation as well.* □

In all what follows we will assume that circuits have a single output node or that, if this is not the case, we have singled out one of them. In this way, we will be only interested in associated functions of the form $f_{\mathcal{C}} : \mathbb{R}^n \to \mathbb{R}$. In addition, we will write

$$S_{\mathcal{C}} := \{x \in \mathbb{R}^n \mid f_{\mathcal{C}}(x) \geq 0\}.$$

Algebraic circuits are easily encoded as points in $\mathcal{I}$ and can therefore be passed as input data to finite-precision BSS machines. We will define two decision problems based on this fact soon enough. But before doing so we want to state a fundamental property of the simulation of the canonical procedure by BSS machines.

**Lemma 2** *There exists a finite-precision BSS machine that, with input a circuit $\mathcal{C}$ with $n$ input nodes and a point $x \in \mathbb{R}^n$ computes $f_{\mathcal{C}}(x)$ following the canonical procedure. The machine does not read the values of $\mathsf{k_{mach}}$ or $\mathsf{u_{mach}}$. The arithmetic cost of this computation is linear in $\mathsf{length}(\mathcal{C})$. Furthermore, for every $\varepsilon \in (0,1)$ the set $\mathsf{Comp}(M, \varepsilon, \mathcal{C}, x)$ of $M$ bijects with the set of possible $\varepsilon$-evaluations of $\mathcal{C}$ on input $x$.*

PROOF. The existence of a machine $M$ evaluating circuits on points via the canonical procedure is clear. Furthermore, we note that all the data management of such $M$ is performed with discrete data and is therefore error-free. The only real number arithmetic performed by $M$ corresponds to the operations of $\mathcal{C}$, and the fact that $M$ follows the canonical procedure means that both $M$ and $\mathcal{C}$ evaluate $f_{\mathcal{C}}$ following the same sequence of arithmetic operations. It follows from this that to each $\varepsilon$-computation of $M$ with input $(\mathcal{C}, x)$ corresponds a $\varepsilon$-evaluation of $\mathcal{C}$ on $x$, and conversely. □

The next decision problem will be essential in the sequel.

**Example 9** Instances for $\mathsf{CircEval}$ are algebraic circuits $\mathcal{C}$ (with input variables $X_1, \ldots, X_n$) together with a point $x \in \mathbb{R}^n$. The problem is to decide whether $x \in S_{\mathcal{C}}$.

To specify a condition number we first define

$$\varrho_{\mathsf{eval}}(\mathcal{C}, x) := \begin{cases} \sup\{\varepsilon < 1 \mid \text{all } \varepsilon\text{-evaluations of } \mathcal{C} \text{ at } x \text{ yield } x \in S_{\mathcal{C}}\} & \text{if } x \in S_{\mathcal{C}} \\ \sup\{\varepsilon < 1 \mid \text{all } \varepsilon\text{-evaluations of } \mathcal{C} \text{ at } x \text{ yield } x \notin S_{\mathcal{C}}\} & \text{otherwise.} \end{cases}$$

We then take as condition number

$$\mu_{\mathsf{eval}}(\mathcal{C}, x) := \max\left\{1, \frac{1}{\varrho_{\mathsf{eval}}(\mathcal{C}, x)}\right\}.$$

In case $(\mathcal{C}, x)$ is syntactically incorrect (e.g., $\mathcal{C}$ is not properly encoded, $x \in \mathbb{R}^s$ with $s \neq n$) we set $(\mathcal{C}, x) \notin \mathsf{CircEval}$ and take $\mu_{\mathsf{eval}}(\mathcal{C}, x) := 1$.

**Proposition 5** *We have* $\mathsf{CircEval} \in \mathrm{P}_{\mathsf{dir}}$.

PROOF. We consider the machine, given by Lemma 2, that with input $(\mathcal{C}, x)$ computes $f_{\mathcal{C}}(x)$ and accepts if the result of this computation is greater than or equal to zero.

The arithmetic cost of this computation is linear in $\mathsf{length}(\mathcal{C})$. Hence, we only need to check that the machine decides correctly as long as its precision is polynomially bounded on $\mathsf{size}(\mathcal{C}, x)$.

For a well-posed input $(\mathcal{C}, x)$ let $\mathsf{u_{mach}} := 2^{-\mathsf{size}(\mathcal{C},x)}$. Then

$$\mathsf{u_{mach}} = \frac{1}{2^{\mathsf{length}(\mathcal{C},x)+\lceil \log \mu(\mathcal{C},x)\rceil}} < \frac{1}{\mu(\mathcal{C}, x)} = \varrho_{\mathsf{eval}}(\mathcal{C}, x)$$

and the definition of $\varrho_{\mathsf{eval}}(\mathcal{C}, x)$ ensures that every $\mathsf{u_{mach}}$-evaluation of $\mathcal{C}$ at $x$ yields $f_{\mathcal{C}(x)} \geq 0$ if and only if $x \in S_{\mathcal{C}}$. That is, the $\mathsf{u_{mach}}$-evaluation yields the same sign ($\geq 0$ or $< 0$) than the infinite-precision evaluation. In other words, this $\mathsf{u_{mach}}$-evaluation correctly decides the input. Since the choice of $\mathsf{u_{mach}}$ is equivalent to set $\mathsf{k_{mach}} := \mathsf{size}(\mathcal{C}, x)$, we are done. $\square$

The (discrete) class $\mathsf{P}$ is included in $\mathrm{P}_{\mathsf{dir}}$ via the simulation of Turing machines mentioned in Remark 6 (we take condition one for all inputs in a discrete problem). We will see in §4.5 that $\mathrm{P}_{\mathsf{dir}}$ is also closely related to $\mathrm{P}_{\mathbb{R}}$.

### 4.4 Variable-precision: the class $\mathrm{P}_{\mathsf{iter}}$

The definition of $\mathrm{P}_{\mathsf{ro}}$ requires that, for a given input $(u, x)$, the machine correctly decides the membership of $(u, x)$ to $S$ as soon as $\mathsf{k_{mach}}$ is large enough. But it does not impose any specific output otherwise. If the precision is insufficient, an output in $\{\mathtt{Yes}, \mathtt{No}\}$ may be wrong. In general, this possibility cannot be ruled out a priori as we do not know the condition $\boldsymbol{\mu}(u, x)$ of the input and, consequently, cannot estimate whether the available precision is sufficient or not for the input at hand. The same can be said of $\mathrm{P}_{\mathsf{dir}}$.

For some decision problems, however, this uncertainty can be avoided. This is the case when there is a simple procedure to guarantee, given $(u, x)$ and $\mathsf{k_{mach}}$, that the computed output is correct. Such a situation would therefore allow three possible outputs: `Yes`, `No`, and `Unsure`. The first two being guaranteed correct, and the last meaning "I need more resources to decide this input." Availability of such a procedure naturally introduces the consideration of *variable precision* algorithms.

These are iterative algorithms which can modify the value of their precision $\mathsf{k_{mach}}$ and always return the correct output (`Yes` or `No`) to the decision question. They adaptively increase their precision (in theory re-reading their input each time they do so) and only halt when the current precision, together with the computations done, guarantee a correct answer. That is, their form typically follows the following general scheme:

```
input (u, x)
initialize k_mach
repeat
    attempt to decide (u, x) and halt if the outcome        (GS)
        is either Yes or No
```

`if the outcome is Unsure then increase` $\mathsf{k}_{\mathsf{mach}}$

One can therefore define a subclass $\mathrm{P}_{\mathsf{iter}}$ of $\mathrm{P}_{\mathsf{ro}}$ that contains the problems for which such an error-free algorithm exists (with the appropriate polynomial cost bounds).

**Definition 8** A decision problem $(S, \boldsymbol{\mu})$ belongs to $\mathrm{P}_{\mathsf{iter}}$ (*iterative polynomial cost*) when there exists a BSS machine $M$ in $\mathrm{P}_{\mathsf{ro}}$ such that for all $(u,x)$ with $\boldsymbol{\mu}(u,x) < \infty$:

**(i)** all computations of $M$ return an element in $\{\mathtt{Yes}, \mathtt{No}, \mathtt{Unsure}\}$ and in the first two cases this output is correct, and

**(ii)** there exist $C, p > 0$ such that if

$$\mathsf{k}_{\mathsf{mach}} \geq C\,\mathsf{size}(u,x)^p$$

then all computations of $M$ return an element in $\{\mathtt{Yes}, \mathtt{No}\}$.

We refer to the scheme (GS) coupled with $M$ as a *machine in* $\mathrm{P}_{\mathsf{iter}}$.

**Remark 12** Unlike machines in $\mathrm{P}_{\mathsf{dir}}$, the halting time of a machine in $\mathrm{P}_{\mathsf{iter}}$ is not bounded. It increases with the input size and may be infinite (i.e., the machine may loop forever) for ill-posed inputs. On the other hand, outputs of $\mathrm{P}_{\mathsf{iter}}$ machines are always correct.

We have already described (without saying so) a problem in $\mathrm{P}_{\mathsf{iter}}$.

**Example 5 (continued)** It was shown in [21] (see also [10, Section 9.4]) that the feasibility of a system $Ay \geq 0, y \neq 0$, can be decided using an iterative algorithm that follows the general scheme (GS). The algorithm carries out

$$\mathcal{O}\big(\sqrt{n}(\log n + \log \mathscr{C}(A))\big)$$

iterations, each of them performing $\mathcal{O}(n^3)$ arithmetic operations. The value of $\mathsf{u}_{\mathsf{mach}}$ is refined at each iteration and the finest value used by the algorithm satisfies

$$\mathsf{u}_{\mathsf{mach}} = \frac{1}{\mathcal{O}\big(n^{12}\mathscr{C}(A)^2\big)}.$$

These bounds show the problem is in $\mathrm{P}_{\mathsf{iter}}$.

The following result is trivial.

**Proposition 6** *We have* $\mathrm{P}_{\mathsf{iter}} \subset \mathrm{P}_{\mathsf{ro}}$. □

### 4.5 Some remarks on infinite precision

Most of the literature in numerical analysis describes algorithms in a context of infinite precision. Finite precision analyses are tedious and, more often than not, avoided. It is therefore worth to ponder on what the classes $\mathrm{P}_{\mathsf{dir}}$ and $\mathrm{P}_{\mathsf{iter}}$ become under the presence of infinite precision. For this, one replaces in the definition of these two classes and if needed, the bound for $\mathsf{k}_{\mathsf{mach}}$ in the bound for the arithmetic cost (so that the later are in terms of input's size) and then disregards the requirement on $\mathsf{k}_{\mathsf{mach}}$ altogether. Because the assumption of infinite precision puts us on the standard BSS realm, the classes thus obtained will be either new or already existing classes in the standard theory.

By definition, the arithmetic cost of machines in $\mathrm{P}_{\mathsf{dir}}$ is independent on the precision at hand or the input's condition. Also, since we are assuming infinite precision, the issue of how large needs $\mathsf{k}_{\mathsf{mach}}$ to be to guarantee a correct answer becomes irrelevant. All in all, condition plays no role and the following straightforward result shows that, under the presence of infinite precision, the class $\mathrm{P}_{\mathsf{dir}}$ is, essentially, $\mathrm{P}_{\mathbb{R}}$.

**Proposition 7** *Let $S \subset \mathcal{I}$. If $(S, \boldsymbol{\mu}) \in \mathrm{P}_{\mathsf{dir}}$ then $S \in \mathrm{P}_{\mathbb{R}}$. Conversely, if $S \in \mathrm{P}_{\mathbb{R}}$ then $(S, \mu_\infty) \in \mathrm{P}_{\mathsf{dir}}$. Here $\mu_\infty$ is the constant function with value $\infty$.* □

Consider now a computation in $\mathrm{P}_{\mathsf{dir}}$ endowed with infinite precision. Because the computations are error-free, the only role played by $\mathsf{k}_{\mathsf{mach}}$ is in allowing more computing time at each iteration. And part (ii) in Definition 8 puts a bound on the total running time in terms of the input's condition (or size). The complexity class emerging captures many algorithms described in the literature.

We say that a problem $(S, \boldsymbol{\mu})$ is in $\mathrm{P}_\infty$ when there exists a (standard) BSS machine deciding $(S, \boldsymbol{\mu})$ whose running time on input $(u, x) \in \mathcal{I}$ is bounded by a polynomial in $\mathsf{size}(u, x)$.

This is a new class in the standard BSS setting, the first one (to the best of our knowledge) to consider condition as a complexity parameter. One clearly has $\mathrm{P}_{\mathsf{iter}} \subset \mathrm{P}_\infty$.

## 5 Nondeterministic Polynomial Cost

Problems in (all versions of) NP are sets $S$ for which membership of an element $x$ to $S$ can be established through an "easy" proof $y$. All definitions of the discrete class $\mathsf{NP}$ translate that $y$ is easy by imposing that the length of $y$ must be polynomially bounded in the length of $x$ (in addition of the fact that one can verify that $y$ is a proof of $x \in S$ in polynomial time in the length of $(x, y)$). Similarly for the class $\mathrm{NP}_{\mathbb{R}}$.

The finite-precision context introduces additional considerations. Not only the length of a proof $y$ will have to be appropriately bounded, it is clear that the condition of the pair $(x, y)$ will have to be so as well. The class $\mathrm{NP}_{\mathsf{ro}}^{\mathsf{U}}$ is obtained by imposing these restrictions.

Even more so, we might want the magnitude of a proof $y$, understood as how large or small can the absolute value of the components of $y$ be, to be bounded as well. This gives rise to the class $\mathrm{NP}_{\mathsf{ro}}^{\mathsf{B}}$.

## 5.1 The class $\mathrm{NP}_{\mathsf{ro}}^{\mathsf{U}}$

**Definition 9** A decision problem $(W, \boldsymbol{\mu}_W)$ belongs to $\mathrm{NP}_{\mathsf{ro}}^{\mathsf{B}}$ (*non-deterministic roundoff polynomial cost*) when there exist a decision problem $(B, \boldsymbol{\mu}_B)$, a finite-precision BSS machine $M$ deciding $(B, \boldsymbol{\mu}_B)$ in $\mathrm{P}_{\mathsf{ro}}$, and polynomials $p, Q$, such that for $(u, x) \in \mathcal{I}$,

**(i)** if $(u, x) \in W$ then there exists $y^* \in \mathbb{R}^m$, such that $(u, x, y^*) \in B$, and $\log \boldsymbol{\mu}_B(u, x, y^*) \leq Q(\log \boldsymbol{\mu}_W(u, x))$, and

**(ii)** if $(u, x) \notin W$ then, for all $y \in \mathbb{R}^m$ we have $(u, x, y) \notin B$ and $\log \boldsymbol{\mu}_B(u, x, y) \leq Q(\log \boldsymbol{\mu}_W(u, x))$.

Here $m = p(\mathsf{length}(u, x))$.

**Example 10** Instances for $\mathsf{CircFeas}$ are algebraic circuits $\mathcal{C}$ (with input variables $Y_1, \ldots, Y_m$). The problem is to decide whether there exists $y \in \mathbb{R}^m$ such that $y \in S_{\mathcal{C}}$ (in which case, we say that $\mathcal{C}$ is *feasible*). We take as condition number

$$\mu_{\mathsf{feas}}(\mathcal{C}) := \max\left\{1, \frac{1}{\varrho_{\mathsf{feas}}(\mathcal{C})}\right\}$$

where

$$\varrho_{\mathsf{feas}}(\mathcal{C}) := \begin{cases} \sup_{y \in S_{\mathcal{C}}} \varrho_{\mathsf{eval}}(\mathcal{C}, y) & \text{if } \mathcal{C} \text{ is feasible,} \\ \inf_{y \in \mathbb{R}^m} \varrho_{\mathsf{eval}}(\mathcal{C}, y) & \text{otherwise.} \end{cases}$$

Note that in the feasible case, $\mu_{\mathsf{feas}}(\mathcal{C})$ is the condition of its best conditioned solution, and in the infeasible case, it is the condition of the worst conditioned point in $\mathbb{R}^m$.

**Proposition 8** *We have* $\mathsf{CircFeas} \in \mathrm{NP}_{\mathsf{ro}}^{\mathsf{U}}$.

Proof. We need to exhibit a problem $(B, \boldsymbol{\mu}_B)$ and a machine $M$ as in Definition 9. For $(B, \boldsymbol{\mu}_B)$ we take the problem $\mathsf{CircEval}$, which we know is in $\mathrm{P}_{\mathsf{ro}}$ by Proposition 5. For $M$ we take the machine on the proof of that proposition (which actually shows that $\mathsf{CircEval} \in \mathrm{P}_{\mathsf{dir}}$).

Let $\mathcal{C}$ be a circuit having $m$ input variables. If $\mathcal{C} \in \mathsf{CircFeas}$ then there exist points $y \in S_{\mathcal{C}}$. Choose $y^*$ among those, that additionally satisfies[1]

$$\varrho_{\mathsf{feas}}(\mathcal{C}) = \varrho_{\mathsf{eval}}(\mathcal{C}, y^*).$$

Since $m \leq \mathsf{length}(\mathcal{C})$ the first requirement in Definition 9(i) is trivially true. In addition, $\mu_{\mathsf{feas}}(\mathcal{C}) = \mu_{\mathsf{eval}}(\mathcal{C}, y^*)$, which implies $\mu_{\mathsf{eval}}(\mathcal{C}, y^*) \leq \mu_{\mathsf{feas}}(\mathcal{C})$ and hence the second requirement.

Now assume that $\mathcal{C} \notin \mathsf{CircFeas}$. Then, for all $y \in \mathbb{R}^m$, $(\mathcal{C}, y) \notin B$. The second requirement in Definition 9(ii) is immediate since the definition of $\mu_{\mathsf{feas}}(\mathcal{C})$ implies that, in the infeasible case, $\varrho_{\mathsf{feas}}(\mathcal{C}) \leq \varrho_{\mathsf{eval}}(\mathcal{C}, y)$ for all $y \in \mathbb{R}^m$, and hence, that $\mu_{\mathsf{eval}}(\mathcal{C}, y) \leq \mu_{\mathsf{feas}}(\mathcal{C})$. □

**Proposition 9** *We have* $\mathrm{P}_{\mathsf{ro}} \subset \mathrm{NP}^{\mathsf{U}}_{\mathsf{ro}}$.

PROOF. Let $(W, \boldsymbol{\mu}_W) \in \mathrm{P}_{\mathsf{ro}}$. Let $(B, \boldsymbol{\mu}_B)$ where $B = W \times \mathbb{R}$ and $\boldsymbol{\mu}_B(u, x, y) = \boldsymbol{\mu}_W(u, x)$. Let $M'$ be any machine deciding $(W, \boldsymbol{\mu}_W)$ in $\mathrm{P}_{\mathsf{ro}}$ and $M$ the machine that with input $(u, x, y)$ runs $M'$ on $(u, x)$. Then $M$ shows that $(B, \boldsymbol{\mu}_B) \in \mathrm{P}_{\mathsf{ro}}$ and the pair $((B, \boldsymbol{\mu}_B), M)$ establishes that $(W, \boldsymbol{\mu}_W) \in \mathrm{NP}^{\mathsf{U}}_{\mathsf{ro}}$ (with $p = 1$ and $Q(n) = n$). □

**Definition 10** A $\mathrm{P}_{\mathsf{ro}}$-*reduction* from $(W, \boldsymbol{\mu}_W)$ to $(S, \boldsymbol{\mu}_S)$ is a finite-precision machine $\overline{M}$ which, given a point $(u, x) \in \mathcal{I}$ and a number $k \in \mathbb{N}$, performs a discrete computation and returns a pair $(v, z) \in \mathcal{I}$ with $\mathsf{ar_cost}_{\overline{M}}(u, x)$ polynomially bounded on $\mathsf{length}(u, x)$ and $k$.

In addition, we require the existence of some $D, p > 0$ such that for all $k \geq D\,\mathsf{size}(u, x)^p$ one has

**(i)** $(u, x) \in W \iff (v, z) \in S$, and

**(ii)** $\log \boldsymbol{\mu}_S(v, z)$ is polynomially bounded in $\mathsf{size}_W(u, x)$.

If all of the above holds, we write $(W, \boldsymbol{\mu}_W) \preceq_{\mathsf{ro}} (S, \boldsymbol{\mu}_S)$.

The notion of $\mathrm{P}_{\mathsf{ro}}$-reduction is tailored to capture the basic property any polynomial-time reduction must have (Proposition 10 below) together with the peculiarities of finite-precision computations.

**Proposition 10** *If* $(W, \boldsymbol{\mu}_W) \preceq_{\mathsf{ro}} (S, \boldsymbol{\mu}_S)$ *and* $(S, \boldsymbol{\mu}_S) \in \mathrm{P}_{\mathsf{ro}}$ *then* $(W, \boldsymbol{\mu}_W) \in \mathrm{P}_{\mathsf{ro}}$.

[1] Such a $y^*$ may not exists since the sup in the definition of $\varrho_{\mathsf{feas}}(\mathcal{C})$ may not be attained. As the modification to the proof in such a case is straightforward, for the sake of clarity and conciseness, we will write the proof for the case that $y^*$ exists.

PROOF. Let $\overline{M}$ be as in the definition above and $N_S$ be a finite-precision machine solving $S$ in $\mathrm{P}_{\mathsf{ro}}$.

By hypothesis, the arithmetic cost of $\overline{M}$ on input $(u,x)$ is bounded by a polynomial in $\mathsf{length}(u,x)$ and $k$, and therefore, such a bound also holds for $\mathsf{length}(v,z)$ (since the cost of writing the output is smaller than the total arithmetic cost). That is, there exist constants $A,t\in\mathbb{N}$, such that

$$\mathsf{length}(v,z)\leq A(\mathsf{length}(u,x)k)^t. \tag{12}$$

Also, there exists $D,E,p,s\in\mathbb{N}$ such that for any input $(u,x)$, if

$$k\geq D\,\mathsf{size}(u,x)^p \tag{13}$$

then

$$\log\boldsymbol{\mu}_S(v,z)\leq E\,\mathsf{size}_W(u,x)^s \quad\text{and}\quad (v,z)\in S\iff(u,x)\in W. \tag{14}$$

Finally, we know that for some $C,q\in\mathbb{N}$, and for any input $(v,z)$, if

$$\mathsf{k_{mach}}(N_S)\geq C\,\mathsf{size}(v,z)^q \tag{15}$$

then $N_S$ correctly decides whether $(v,z)\in S$.

Let $\mathcal{M}$ be the machine given by the following code:

> `input` $(u,x)$
> `compute` $k:=\left\lfloor\mathsf{k}_{\mathsf{mach}}^{\frac{1}{2qt}}\right\rfloor$
> `run` $\overline{M}$ `on input` $((u,x),k)$`; let` $(v,z)$ `be the returned point`
> `run` $N_S$ `on input` $(v,z)$ `and accept iff` $N_S$ `accepts`

We will prove that this machine decides $(W,\boldsymbol{\mu}_W)$ in $\mathrm{P}_{\mathsf{ro}}$. We begin by observing that, by our choice of $k$, we have

$$k^{2qt}\leq\mathsf{k_{mach}}\leq(k+1)^{2qt}. \tag{16}$$

We first deal with the complexity. We know that the arithmetic cost of $\overline{M}$ is polynomial in $\mathsf{length}(u,x)$ and $k$. Since the latter is bounded by $\mathsf{k_{mach}}(\mathcal{M})$ we are done with the cost of $\overline{M}$. And the cost of $N_S$ is also polynomially bounded in $\mathsf{length}(u,x)$ and $\mathsf{k_{mach}}(\mathcal{M})$ since such a bound holds on $\mathsf{length}(v,z)$ and $\mathsf{k_{mach}}(\mathcal{M})$ and, by (12), the first term is polynomially bounded on $\mathsf{length}(u,x)$ and $k$.

We next deal with the precision needed. Set

$$\mathsf{k_{mach}}:=\left\lceil\max\left\{C^2(2A)^{2q}\mathsf{size}(u,x)^{2qt},C\,(2E)^q(\mathsf{size}(u,x){+}1)^{qs},(D\,\mathsf{size}(u,x)^p{+}1)^{2qt}\right\}\right\rceil.$$

This value of $\mathsf{k_{mach}}$ is clearly polynomially bounded on $\mathsf{size}(u,x)$. Now, because of the last term within the brackets (and (16)) we have

$$k\geq\mathsf{k}_{\mathsf{mach}}^{\frac{1}{2qt}}-1\geq D\,\mathsf{size}(u,x)^p,$$

i.e., (13) holds. It follows that the simulation of $\overline{M}$ on input $((u,x),k)$ returns $(v,z)$ satisfying (14).

In addition, we have

$$\mathsf{size}(v,z) \leq 2\max\{\log \boldsymbol{\mu}_S(v,z)+1, \mathsf{length}(v,z)\}.$$

We divide by cases.

1) Assume first that $\mathsf{size}(v,z) \leq 2\,\mathsf{length}(v,z)$. Then, using (12)

$$\begin{aligned} C\,\mathsf{size}(v,z)^q &\leq C(2A)^q \mathsf{length}(u,x)^{qt} k^{qt} \\ &\leq C(2A)^q \mathsf{size}(u,x)^{qt} \sqrt{\mathsf{k}_{\mathsf{mach}}} \\ &\leq \mathsf{k}_{\mathsf{mach}} \end{aligned}$$

the last inequality since $\mathsf{k}_{\mathsf{mach}} \geq C^2(2A)^{2q}\mathsf{size}(u,x)^{2qt}$.

2) Assume instead that $\mathsf{size}(v,z) \leq 2(\log \boldsymbol{\mu}_S(v,z)+1)$. Then, and here we use the first statement in (14),

$$\begin{aligned} C\,\mathsf{size}(v,z)^q &\leq C\,2^q(\log \boldsymbol{\mu}_S(v,z)+1)^q \\ &\leq C\,2^q(E(\mathsf{size}_W(u,x))^s+1)^q \\ &\leq C\,(2E)^q(\mathsf{size}(u,x)+1)^{qs} \\ &\leq \mathsf{k}_{\mathsf{mach}}. \end{aligned}$$

In both cases the precision on the simulation of $N$ on input $(v,z)$ satisfies (15) and therefore we have both that $(v,z) \in S$ iff $(u,x) \in W$ and that $N_S$ correctly decides whether $(v,z) \in S$. It follows that $\mathcal{M}$ correctly decides whether $(u,x) \in W$. $\square$

**Definition 11** We say that a decision problem $(S,\boldsymbol{\mu}_S)$ is $\mathrm{NP}^{\mathsf{U}}_{\mathsf{ro}}$-*hard* when for any problem $(W,\boldsymbol{\mu}_W) \in \mathrm{NP}^{\mathsf{U}}_{\mathsf{ro}}$ we have $(W,\boldsymbol{\mu}_W) \preceq_{\mathsf{ro}} (S,\boldsymbol{\mu}_S)$. We say that it is $\mathrm{NP}^{\mathsf{U}}_{\mathsf{ro}}$-*complete* when it is $\mathrm{NP}^{\mathsf{U}}_{\mathsf{ro}}$-hard and it belongs to $\mathrm{NP}^{\mathsf{U}}_{\mathsf{ro}}$.

**Theorem 1** *The problem* CircFeas *is* $\mathrm{NP}^{\mathsf{U}}_{\mathsf{ro}}$*-complete.*

PROOF. We have already seen in Proposition 8 that CircFeas is in $\mathrm{NP}^{\mathsf{B}}_{\mathsf{ro}}$. The hardness of CircFeas applies arguments that have been used once and again, adapted to our context.

Consider a problem $(W,\boldsymbol{\mu}_W)$ and a pair $((B,\boldsymbol{\mu}_B),M)$ as in Definition 9 certifying this problem in $\mathrm{NP}^{\mathsf{B}}_{\mathsf{ro}}$. Let $p,q,Q$ and $m$ be as in that definition. Also, let $\mathsf{Arith}_M$ and $\mathsf{Prec}_M$ be the functions bounding the arithmetic cost and necessary precision of $M$, as in Definition 5. Without loss of generality, we will assume that $\mathsf{Prec}_M$ is monotonically increasing.

Next fix an input $(u,x) \in \mathcal{I}$, $k \in \mathbb{N}$ and let $\ell := \mathsf{length}(u,x)$, $m = p(\ell)$, and $T := \mathsf{Arith}_M(\ell+m,k)$. Then, for all $y \in \mathbb{R}^m$, all computations of $M$ with input $(u,x,y)$ and precision $\mathsf{k}_{\mathsf{mach}} = k$ halt and return an output in $\{\mathtt{Yes},\mathtt{No}\}$ within $T$ steps. One can construct a decision circuit $\mathcal{C}$ of depth $T$ in the variables $Y_1, \ldots, Y_m$ (having the numbers $x_1, \ldots, x_n$ associated to constant nodes) which replicates these computations. More precisely, $\mathcal{C}$ satisfies the following conditions:

**(a)** The number of nodes of $\mathcal{C}$ is polynomial in $\ell$ and $k$.

**(b)** For all $y \in \mathbb{R}^m$, there exist accepting $\mathsf{u_{mach}}$-computations of $M$ with input $(u, x, y)$ iff there exist $\mathsf{u_{mach}}$-evaluations of $\mathcal{C}$ at $y$ yielding $y \in S_{\mathcal{C}}$. Similarly for rejecting computations and $y \notin S_{\mathcal{C}}$.

**(c)** The circuit $\mathcal{C}$ is computed with cost polynomial in $\ell$ and $k$ and this computation is discrete.

The construction of this circuit is given with details in [23] so we won't repeat these details here. We may nonetheless summarize the main idea (but, note, this requires familiarity with BSS machines as presented in [6] from the part of the reader).

At any time during the computation of $M$ the internal state of the machine can be described by the current node $\eta \in \{1, \ldots, N\}$ of $M$ together with an element in its state space $\mathbb{N} \times \mathbb{N} \times \mathbb{R}^\infty$. If the computation performs $T$ steps then the values of the first two components are themselves bounded by $T$ and the only components of $\mathbb{R}^\infty$ that ever play a role in it are the first $T$. It follows that the relevant variable taking values during the computation are the following:

$i_t, j_t \in \{0, \ldots, T\}$: for the values of the two integer components of the state space at time $t = 0, \ldots, T$,

$w_t \in \{1, \ldots, N\}$: for the value of the current node at time $t = 0, \ldots, T$,

$z_{s,t} \in \mathbb{R}$: for the value of the $s$th component of the state space at time $t$, $s, t = 0, \ldots, T$.

The values of these variables at time $t = 0$ are given by the initialization of the machine. For $t \geq 1$, these values depend on the values of a few (at most 8, see [23]) variables at time $t-1$. In addition, this dependence is simple in the sense that it can be computed by circuits $\mathcal{I}_t$, $\mathcal{J}_t$, $\mathcal{W}_t$, and $\mathcal{Z}_{s,t}$, each of them with a small, bounded number of nodes. The circuit $\mathcal{C}$ above is obtained by appropriately connecting these four families plus a series of constant nodes corresponding to the constants of $M$, the components of $x$, and the numbers $\mathsf{u_{mach}}$ and $\mathsf{k_{mach}}$. A further small subcircuit is required, which returns 1 if the computation of $M$ accepts and returns $-1$ otherwise. It is straightforward to see that it satisfies conditions (a), (b), and (c) above.

The $\mathrm{P_{ro}}$-reduction is given by the machine $\overline{M}$ which, given $(u, x)$, returns $\mathcal{C}$.

We will prove that this is indeed a reduction. The first condition in Definition 10, the fact that $\mathsf{ar_cost}_{\overline{M}}(u, x)$ is polynomially bounded in $\mathsf{length}(u, x)$ and $k$, is just property (c) above. We therefore focus on the other two conditions in Definition 10, which require to find an appropriate lower bound for $k$.

Let $Q$ be the polynomial in Definition 9 and

$$Y := \{y \in \mathbb{R}^m \mid \log \boldsymbol{\mu}_B(u,x,y) \le Q(\log \boldsymbol{\mu}_W(u,x))\}.$$

Then, for all $y \in Y$,

$$\begin{aligned} \mathsf{size}_B(u,x,y) &= \mathsf{length}(u,x) + m + \lceil \log \boldsymbol{\mu}_B(u,x,y) \rceil \\ &\le \ell + p(\ell) + \lceil Q(\log \boldsymbol{\mu}_W(u,x)) \rceil \le R(\mathsf{size}_W(u,x)) \end{aligned} \tag{17}$$

for a polynomial $R$.

Take any $k$ satisfying

$$k \ge k_0 := \mathsf{Prec}_M(R(\mathsf{size}(u,x))).$$

Note that $k_0$ is polynomially bounded in $\mathsf{size}(u,x)$. We claim that requirements (i) and (ii) in Definition 10 hold for these $k$.

To prove this claim, set the precision of $M$ to be $\mathsf{u_{mach}} = 2^{-k_0}$ (and, accordingly, $\mathsf{k_{mach}} := k_0$). The monotonicity of $\mathsf{Prec}_M$ together with (17) imply that, for all $y \in Y$,

$$\mathsf{k_{mach}} = \mathsf{Prec}_M(R(\mathsf{size}_W(u,x))) \ge \mathsf{Prec}_M(\mathsf{size}_B(u,x,y)). \tag{18}$$

We now divide by cases.

**Case I:** $(u,x) \in W$. In this case, there exists $y^* \in Y$ such that $(u,x,y^*) \in B$. Since $y^* \in Y$, inequality (18) holds true for $y^*$. Therefore, all $\mathsf{u_{mach}}$-computations of $M$ with input $(u,x,y^*)$ halt and accept. It follows from property (b) above that all $\mathsf{u_{mach}}$-evaluations of $\mathcal{C}$ at $y^*$ return $f_{\mathcal{C}}(y^*) \ge 0$. Since this occurs, in particular, for the exact evaluation, we deduce that $\mathcal{C} \in \mathsf{CircFeas}$. This proves (i) in Definition 10. Furthermore, from the fact that all $\mathsf{u_{mach}}$-evaluations of $\mathcal{C}$ with input $y^*$ yield $y \in S_{\mathcal{C}}$ we deduce that $\varrho_{\mathsf{eval}}(\mathcal{C}, y^*) \ge \mathsf{u_{mach}}$ and therefore, that

$$\varrho_{\mathsf{feas}}(\mathcal{C}) \ge \varrho_{\mathsf{eval}}(\mathcal{C}, y^*) \ge \mathsf{u_{mach}}.$$

It follows that $\mu_{\mathsf{feas}}(\mathcal{C}) \le \frac{1}{\mathsf{u_{mach}}} = 2^{k_0}$ and therefore, that

$$\log \mu_{\mathsf{feas}}(\mathcal{C}) \le k_0 = \mathsf{Prec}_M(R(\mathsf{size}_W(u,x))).$$

This bound is polynomial in $\mathsf{size}_W(u,x)$, as we wanted.

**Case II:** $(u,x) \notin W$. In this case, for every $y \in \mathbb{R}^m$, we have $(u,x,y) \notin B$ and $y \in Y$. Again, for all $y \in \mathbb{R}^m$, inequality (18) holds and we deduce that all $\mathsf{u_{mach}}$-computations of $M$ reject $(u,x,y)$. It follows from property (b) that the same result occurs for all $\mathsf{u_{mach}}$-evaluations of $\mathcal{C}$ at $y$. And since these evaluations include the exact one, we deduce that $y \notin S_{\mathcal{C}}$. This is true for all $y \in \mathbb{R}^m$. Therefore, $\mathcal{C}$ is infeasible. This proves condition (i) in Definition 10.

To prove condition (ii), we use again the fact that, for all $y \in \mathbb{R}^m$, all $\mathsf{u_{mach}}$-evaluations of $\mathcal{C}$ at $y$ yield the same output ($y \notin S_{\mathcal{C}}$). This implies that, for all $y \in \mathbb{R}^m$,

$$\varrho_{\mathsf{eval}}(\mathcal{C}, y) \geq \mathsf{u_{mach}} = 2^{-k_0}$$

and therefore, that $\varrho_{\mathsf{feas}}(\mathcal{C}) \geq 2^{-k_0}$. Consequently, we have $\mu_{\mathsf{feas}}(\mathcal{C}) \leq 2^{k_0}$, and we conclude as in Case I. □

The following result is an immediate consequence of Proposition 10 and Theorem 1.

**Corollary 1** *We have* $\mathrm{P}_{\mathsf{ro}} = \mathrm{NP}^{\mathsf{U}}_{\mathsf{ro}} \iff \mathsf{CircFeas} \in \mathrm{P}_{\mathsf{ro}}$. □

**Remark 13** The construction of a circuit as in the proof of Theorem 3 has been done in many situations: for discrete computations it is the basis of Ladner's proof [40] of the $\mathsf{P}$-completeness of the circuit evaluation problem (in this case, condition (c) in the proof is strengthened to require that the computation of $\mathcal{C}$ can be done with fewer resources, usually logarithmic space or polylogarithmic parallel time), in the BSS model is the basis of a similar result over the reals [23], and even in the additive BSS model (where no multiplications are allowed) it is the basis of the proofs of some completeness results [18, 39]. In fact, the universality of this construction has prompted Bruno Poizat [44] to define $\mathsf{P}$ over an arbitrary structure as the class of sets decidable by families of circuits (with nodes appropriate for the structure) that can be constructed in polynomial time by a (standard) Turing machine.

**Open Question 1** The main open question in this development is, as one can expect, to decide whether $\mathrm{P}_{\mathsf{ro}} = \mathrm{NP}^{\mathsf{U}}_{\mathsf{ro}}$. As usual, we believe this is not the case.

### 5.2 The class $\mathrm{NP}^{\mathsf{B}}_{\mathsf{ro}}$

To define this class we must first provide a formal definition of the notion of magnitude.

Given $k \in \mathbb{N}$ we consider the set $F_k$ composed of 0 plus all numbers of the form (1) with

$$\beta = 2, \qquad t = k+1, \quad \text{and} \quad -2^k + 1 \leq e \leq 2^{(k+1)} - 1.$$

By construction (recall (2)) $F_k$ is a floating-point system whose elements $y \neq 0$ satisfy

$$2^{-2^k} \leq |y| \leq 2^{2^{k+1}-1}(1 - 2^{-k-1}) \tag{19}$$

and these upper and lower bounds are attained for some elements in $F_k$. Recall, the real intervals defined by (19), along with $\{0\}$, define the range of $F_k$.

For $x \in \mathbb{R}$ we define the *magnitude* of $x$ to be

$$\mathsf{mgt}(x) := \min\{k \geq 1 \mid x \in \mathsf{Range}(F_k)\},$$

and for $x \in \mathbb{R}^n$, $\mathsf{mgt}(x) := \max_{i \leq n} \mathsf{mgt}(x_i)$.

**Definition 12** A decision problem $(W, \boldsymbol{\mu}_W)$ belongs to $\mathrm{NP}^{\mathsf{B}}_{\mathsf{ro}}$ (*bounded non-deterministic roundoff polynomial cost*) when there exist a decision problem $(B, \boldsymbol{\mu}_B)$, a finite-precision BSS machine $M$ deciding $(B, \boldsymbol{\mu}_B)$ in $\mathrm{P}_{\mathsf{ro}}$, and polynomials $p, q, Q$, such that for $(u, x) \in \mathcal{I}$,

**(i)** if $(u, x) \in W$ then there exists $y^* \in \mathbb{R}^m$, such that $(u, x, y^*) \in B$, $\log \boldsymbol{\mu}_B(u, x, y^*) \leq Q(\log \boldsymbol{\mu}_W(u, x))$, and $\mathsf{mgt}(y^*) \leq q(\mathsf{size}_W(u, x))$, and

**(ii)** if $(u, x) \notin W$ then, for all $y \in \mathbb{R}^m$ we have $(u, x, y) \notin B$ and $\log \boldsymbol{\mu}_B(u, x, y) \leq Q(\log \boldsymbol{\mu}_W(u, x))$.

Here $m = p(\mathsf{length}(u, x))$.

**Example 11** Instances for $\mathsf{CircBFeas}$ are algebraic circuits $\mathcal{C}$ (with input variables $Y_1, \ldots, Y_m$). The problem is to decide whether there exists $y \in \mathbb{R}^m$ such that $y \in S_{\mathcal{C}}$. What makes this problem different from $\mathsf{CircFeas}$ is its condition number. Here we take

$$\mu_{\mathsf{Bfeas}}(\mathcal{C}) := \max\left\{1, \frac{1}{\varrho_{\mathsf{Bfeas}}(\mathcal{C})}\right\}$$

where

$$\varrho_{\mathsf{Bfeas}}(\mathcal{C}) := \begin{cases} \sup_{y \in S_{\mathcal{C}}} \varrho_{\mathsf{eval}}(\mathcal{C}, y) 2^{-\mathsf{mgt}(y)} & \text{if } \mathcal{C} \text{ is feasible,} \\ \inf_{y \in \mathbb{R}^m} \varrho_{\mathsf{eval}}(\mathcal{C}, y) & \text{otherwise.} \end{cases}$$

Agai, in the feasible case, $\mu_{\mathsf{feas}}(\mathcal{C})$ is the condition of its best conditioned solution but here we take into account the magnitude of the solution so that feasible circuits having only large-magnitude solutions are poorly conditioned.

**Proposition 11** *We have* $\mathsf{CircFeas} \in \mathrm{NP}^{\mathsf{B}}_{\mathsf{ro}}$.

PROOF. The proof is as that of Proposition 11 just that now, in the feasible case, $y^*$ is chosen to satisfy

$$\varrho_{\mathsf{feas}}(\mathcal{C}) = \varrho_{\mathsf{eval}}(\mathcal{C}, y^*) 2^{-\mathsf{mgt}(y^*)}.$$

The two first requirements in Definition 12(i) are shown as in that proposition and for the third we have

$$\mathsf{size}(\mathcal{C}) \geq \log \mu_{\mathsf{feas}}(\mathcal{C}) = |\log \varrho_{\mathsf{feas}}(\mathcal{C})| = |\log \varrho_{\mathsf{eval}}(\mathcal{C}, y^*)| + \mathsf{mgt}(y^*) \geq \mathsf{mgt}(y^*)$$

which shows this requirement (with $q$ being the identity). □

In the same manner, the following result is shown as Proposition 9.

**Proposition 12** *We have* $\mathrm{P}_{\mathsf{ro}} \subset \mathrm{NP}^{\mathsf{B}}_{\mathsf{ro}} \subset \mathrm{NP}^{\mathsf{U}}_{\mathsf{ro}}$. □

The notions of $\mathrm{NP}^{\mathsf{B}}_{\mathsf{ro}}$-hardness and $\mathrm{NP}^{\mathsf{B}}_{\mathsf{ro}}$-completeness are defined as in Definition 11. Our main result here is the following.

**Theorem 2** *The problem* CircBFeas *is* $\mathrm{NP}^{\mathsf{U}}_{\mathsf{ro}}$*-complete.*

Proof. It follows that of Theorem 2 almost word by word. The only difference is in the arguments for Case I. The point $y^* \in Y$ such that $(u, x, y^*) \in B$ now additionally satisfies that $\mathsf{mgt}(y^*) \le q(\mathsf{size}(u, x))$. This does not affect the proof that all $\mathsf{u_{mach}}$-evaluations of $\mathcal{C}$ at $y^*$ return $f_{\mathcal{C}}(y^*) \ge 0$ and hence that $\mathcal{C}$ is feasible and that $\varrho_{\mathsf{eval}}(\mathcal{C}, y^*) \ge \mathsf{u_{mach}}$. But from the latter we now deduce

$$\varrho_{\mathsf{feas}}(\mathcal{C}, y^*) \ge \mathsf{u_{mach}} 2^{-\mathsf{mgt}(y^*)} \ge \mathsf{u_{mach}} 2^{-q(\mathsf{size}(u,x))}.$$

It follows that $\mu_{\mathsf{feas}}(\mathcal{C}) \le \frac{2^{-q(\mathsf{size}(u,x))}}{\mathsf{u_{mach}}}$ and therefore, that

$$\log \mu_{\mathsf{feas}}(\mathcal{C}) \le q(\mathsf{size}(u, x)) + k_0 = q(\mathsf{size}(u, x)) + \mathsf{Prec}_M(R(\mathsf{size}_W(u, x))).$$

This bound is clearly polynomial in $\mathsf{size}_W(u, x)$, as we wanted. □

**Corollary 2** *We have* $\mathrm{P}_{\mathsf{ro}} = \mathrm{NP}^{\mathsf{B}}_{\mathsf{ro}} \iff \mathsf{CircBFeas} \in \mathrm{P}_{\mathsf{ro}}$. □

### 5.3 The classes $\mathrm{NP}^{\mathsf{U}}_{\mathsf{dir}}$ and $\mathrm{NP}^{\mathsf{B}}_{\mathsf{dir}}$

The definitions of $\mathrm{NP}^{\mathsf{U}}_{\mathsf{dir}}$ and $\mathrm{NP}^{\mathsf{B}}_{\mathsf{dir}}$ are the obvious variations of that for $\mathrm{NP}^{\mathsf{U}}_{\mathsf{ro}}$ and $\mathrm{NP}^{\mathsf{B}}_{\mathsf{ro}}$.

**Definition 13** A decision problem $(S, \boldsymbol{\mu}_S)$ belongs to $\mathrm{NP}^{\mathsf{U}}_{\mathsf{dir}}$ (*non-deterministic direct polynomial cost*) when there exist a decision problem $(B, \boldsymbol{\mu}_B)$, a finite-precision BSS machine $M$ deciding $(B, \boldsymbol{\mu}_B)$ in $\mathrm{P}_{\mathsf{dir}}$, and polynomials $p, Q$, satisfying properties (i) and (ii) of Definition 9. Similarly for $\mathrm{NP}^{\mathsf{B}}_{\mathsf{dir}}$ (and Definition 12).

Also, our first examples of problems in $\mathrm{NP}^{\mathsf{U}}_{\mathsf{dir}}$ and $\mathrm{NP}^{\mathsf{B}}_{\mathsf{dir}}$ follow from a quick look at the proof of Proposition 8.

**Proposition 13** *We have* $\mathsf{CircFeas} \in \mathrm{NP}^{\mathsf{U}}_{\mathsf{dir}}$ *and* $\mathsf{CircBFeas} \in \mathrm{NP}^{\mathsf{B}}_{\mathsf{dir}}$. □

The fact that the arithmetic cost of $\mathrm{P}_{\mathsf{dir}}$ machines depends only on the input's length allows for a simpler form of reduction.

**Definition 14** A $\mathsf{P}$-*reduction* from $(W, \boldsymbol{\mu}_W)$ to $(S, \boldsymbol{\mu}_S)$ is a finite-precision machine $\overline{M}$ which, given an input $(u, x) \in \mathcal{I}$, performs a discrete computation and returns a pair $(v, z) \in \mathcal{I}$ satisfying the following:

**(i)** $(u,x) \in W \iff (v,z) \in S$,

**(ii)** $\mathsf{ar_cost}_{\overline{M}}(u,x)$ is polynomially bounded on $\mathsf{length}(u,x)$, and

**(iii)** $\mathsf{size}_S(v,z)$ is polynomial in $\mathsf{size}_W(u,x)$.

If all of the above holds, we write $(W, \boldsymbol{\mu}_W) \preceq_{\mathsf{P}} (S, \boldsymbol{\mu}_S)$.

**Proposition 14** *If* $(W, \boldsymbol{\mu}_W) \preceq_{\mathsf{P}} (S, \boldsymbol{\mu}_S)$ *and* $(S, \boldsymbol{\mu}_S) \in \mathrm{P}_{\mathsf{dir}}$ *then* $(W, \boldsymbol{\mu}_W) \in \mathrm{P}_{\mathsf{dir}}$.

PROOF. It is a simpler version of the proof of Proposition 14. □

Hardness and completeness with respect of P-reductions are defined as in Definition 11.

**Theorem 3** *The problems* CircFeas *and* CircBFeas *are* $\mathrm{NP}^{\mathsf{U}}_{\mathsf{dir}}$*-complete and* $\mathrm{NP}^{\mathsf{B}}_{\mathsf{dir}}$*-complete, respectively, with respect of* P*-reductions.*

PROOF. Again, Proposition 13 shows that CircFeas is in $\mathrm{NP}^{\mathsf{U}}_{\mathsf{dir}}$ and we only need to prove the hardness. The proof is, essentially, contained in that of Theorem 1. Instead of a family of circuits parameterized by $k \in \mathbb{N}$, we deal with only one circuit whose depth is given by a polynomial in $\mathsf{length}(u,x)$. Property (iii) in Definition 14 is clear. Property (i) is shown word by word as in Theorem 1. Finally, for property (ii), the proof of this theorem shows that $\log \mu_{\mathsf{feas}}(\mathcal{C})$ is polynomially bounded in $\log \boldsymbol{\mu}(u,x)$. And since $\mathsf{length}(\mathcal{C})$ is polynomially bounded in $\mathsf{length}(u,x)$, it follows that $\mathsf{size}(\mathcal{C})$ is polynomially bounded in $\mathsf{size}(u,x)$.

Similarly for CircBFeas and $\mathrm{NP}^{\mathsf{B}}_{\mathsf{dir}}$. □

The following result is an immediate consequence of Proposition 14 and Theorem 3.

**Corollary 3** *We have* $\mathrm{P}_{\mathsf{dir}} = \mathrm{NP}^{\mathsf{U}}_{\mathsf{dir}} \iff \mathsf{CircFeas} \in \mathrm{P}_{\mathsf{dir}}$ *and* $\mathrm{P}_{\mathsf{dir}} = \mathrm{NP}^{\mathsf{B}}_{\mathsf{dir}} \iff \mathsf{CircBFeas} \in \mathrm{P}_{\mathsf{dir}}$. □

**Open Question 2** Again, we leave open the truth of the equality $\mathrm{P}_{\mathsf{dir}} = \mathrm{NP}^{\mathsf{U}}_{\mathsf{dir}}$. And again, we believe that equality does not hold.

# 6 Deterministic Bounds for Nondeterministic Cost

## 6.1 Exponential cost

As we mentioned in the Introduction, a crucial property of NP or $\mathrm{NP}_{\mathbb{R}}$ is that they are subclasses of their corresponding exponential time classes. In the case of the reals, it is even known that the inclusion $\mathrm{NP}_{\mathbb{R}} \subset \mathrm{EXP}_{\mathbb{R}}$ is strict [17]. The main result in this section shows a similar property for $\mathrm{NP}^{\mathsf{B}}_{\mathsf{ro}}$. Before stating it, we define the general class $\mathrm{EXP}_{\mathsf{ro}}$ of exponential cost, along with subclasses extending $\mathrm{P}_{\mathsf{dir}}$ and $\mathrm{P}_{\mathsf{iter}}$.

**Definition 15** A decision problem $(S, \boldsymbol{\mu})$ belongs to $\mathrm{EXP}_{\mathsf{ro}}$ (*roundoff exponential cost*) when there exists a finite-precision BSS machine $M$ deciding $S$ with cost $(\mathsf{Arith}, \mathsf{Prec})$ and such that

**(i)** $\mathsf{Prec}$ is bounded by a exponential function, and

**(ii)** the function $\mathsf{Arith}(\mathsf{length}(u,x), \mathsf{Prec}(\mathsf{size}(u,x)))$ is bounded by an exponential in $\mathsf{size}(u,x)$, for all $(u,x) \in \mathcal{I}$.

In both (i) and (ii) by exponential we understand a function of the kind $n \mapsto a^{n^d}$ for some $a > 1$ and $d > 0$.

**Remark 14** What we observed for Definition 5 in Remark 9(i) applies here *mutatis mutandis*. In particular, when $\mathsf{Prec}$ in Definition 15 is polynomially bounded we say that $(S, \boldsymbol{\mu})$ can be solved with *polynomial precision*, and we write $(S, \boldsymbol{\mu}) \in \mathrm{EXP}_{\mathsf{ro}}^{[\mathsf{P}]}$. It is important to note that in this case the dependence of $\mathsf{Arith}$ on $\mathsf{k_{mach}}$ may be exponential.

The classes $\mathrm{EXP}_{\mathsf{dir}}$ and $\mathrm{EXP}_{\mathsf{iter}}$ are defined with the obvious modifications to Definitions 6 and 8. In both cases, if the required precision satisfies $\mathsf{k_{mach}} = (\mathsf{size}(u,x))^{\mathcal{O}(1)}$ we say that the problem can be solved with *polynomial precision*.

**Example 7 (continued)** The main result in [22] shows that the problem mentioned in Example 7 (feasibility of real homogeneous polynomial systems) is in $\mathrm{EXP}_{\mathsf{iter}}$.

**Proposition 15** *The inclusion* $\mathrm{EXP}_{\mathsf{ro}}^{[\mathsf{P}]} \subset \mathrm{EXP}_{\mathsf{ro}}$ *is strict. The class* $\mathrm{EXP}_{\mathsf{dir}}$ *is not included in* $\mathrm{EXP}_{\mathsf{ro}}^{[\mathsf{P}]}$.

Proof. Proposition 1, with $T(n) = 2^n$, $P_1$ a polynomial function, and $P_2(n) = 2^n$, proves the first statement. A closer look at its proof reveals that the machine deciding the set $(B, \boldsymbol{\mu})$ there, with the functions above, is in $\mathrm{EXP}_{\mathsf{dir}}$. The second statement follows. □

The following results are shown as Propositions 10 and 14.

**Proposition 16** *If* $(W, \boldsymbol{\mu}_W) \preceq_{\mathsf{ro}} (S, \boldsymbol{\mu}_S)$ *and* $(S, \boldsymbol{\mu}_S) \in \mathrm{EXP}_{\mathsf{ro}}$ *then* $(W, \boldsymbol{\mu}_W) \in \mathrm{EXP}_{\mathsf{ro}}$. *A similar statement holds for the class* $\mathrm{EXP}_{\mathsf{ro}}^{[\mathsf{P}]}$. □

**Proposition 17** *If* $(W, \boldsymbol{\mu}_W) \preceq_{\mathsf{P}} (S, \boldsymbol{\mu}_S)$ *and* $(S, \boldsymbol{\mu}_S) \in \mathrm{EXP}_{\mathsf{dir}}$ *then* $(W, \boldsymbol{\mu}_W) \in \mathrm{EXP}_{\mathsf{dir}}$. □

## 6.2 Testing grids

The fact that finite-precision computations need to be robust (i.e., they need to result in the same outcome) when the precision is sufficiently large allows to reduce the search for solutions to the circuit feasibility problem to a search over the points of a sufficiently fine grid. We describe here these grids and the cost and accuracy of constructing them.

We begin by recalling the floating-point system $F_k$ defined at the beginning of §5.1 and observing that its unit roundoff is $u_k := 2^{-(k+1)}$.

A point in $F_k$ can be given by $2k+2$ bits ($k+1$ to write down $e$ along with $k+1$ more to write $d_1, \dots, d_t$). We will denote by $y_v$ the element in $F_k$ associated to a point $v \in \{0,1\}^{2k+2}$.

Consider the set $\mathscr{G}_k := F_k^n \subset \mathbb{R}^n$. For any $\bar{v} = (v_1, \dots, v_n) \in \{0,1\}^{n(2k+2)}$ we write $\overline{y_v} = (y_{v_1}, \dots, y_{v_n}) \in \mathscr{G}_k$.

**Proposition 18** *Given $v \in \{0,1\}^{(2k+2)}$ we can compute $y_v \in \mathbb{R}$ with $\mathcal{O}(k)$ arithmetic operations. If $k \geq 5$ and $\mathsf{u_{mach}} \leq 2^{-2k}$, the computed quantity $\widetilde{y_v}$ satisfies*

$$\widetilde{y_v} = y_v(1 + \theta_{2^{k+2}}).$$

*Furthermore, for all $\varepsilon > 0$ we can ensure*

$$\widetilde{y_v} = y_v(1+\delta) \qquad \text{with } |\delta| \leq \varepsilon$$

*if, in addition, $2^k < \frac{\varepsilon}{2}$.*

PROOF. To compute $y_v$ we need to compute both $2^e$ and $m = 0.d_1 d_2 \dots d_t$. Assume, without loss of generality, that $e > 0$. Since $e < 2^{k+1}$, it has a binary decomposition $e = \sum_{j=0}^{k} b_j 2^j$ (here $b_j \in \{0,1\}$). Hence

$$2^e = \prod_{b_j=1} 2^{2^j}. \tag{20}$$

Since $2^{2^{j+1}} = 2^{2^j} \cdot 2^{2^j}$ we can compute the collection of all the $2^{2^j}$ with $k$ multiplications, and with at most $k$ additional multiplications we obtain $2^e$. The cost of computing $2^e$ is therefore $\mathcal{O}(k)$.

Also, the mantissa

$$m = 0.d_1 d_2 \dots d_t = \sum_{i=1}^{t} d_i 2^{-i}$$

can be computed with at most $2t$ operations. For $i > 1$ each $2^{-i}$ is obtained with a single division from $2^{-(i-1)}$ and is added to the partial sum if $d_i = 1$. It follows that the arithmetic cost of computing $y_v$ is $\mathcal{O}(k)$.

We now consider precision issues.

Recall the computation of the mantissa. The quantity $2^{-i}$ is obtained from $2^{-(i-1)}$ with one division. It is easy to show by induction (using Lemma 1) that the computed quantity $\widetilde{2^{-i}}$ is of the form $2^{-i}(1+\theta_{2i-1})$. From this bound, using that

$$\begin{aligned}
&\left(\left(\sum_{i=1}^{j} d_i 2^{-i}\right)(1+\theta_{2(j+1)-1}) + d_{j+1}2^{-(j+1)}(1+\theta_{2(j+1)-1}\right)(1+\theta_1)\\
=\;&\left(\left(\sum_{i=1}^{j+1} d_i 2^{-i}\right)(1+\theta_{2(j+1)-1})\right)(1+\theta_1) \;=\; \sum_{i=1}^{j+1} d_i 2^{-i}(1+\theta_{2(j+2)-1}),
\end{aligned}$$

a further induction argument shows that the computed mantissa is of the form

$$\widetilde{m} = \left(\sum_{i=i}^{k+1} d_i 2^{-i}\right)(1+\theta_{2(k+2)-1}).$$

Recall also the computation of $y_v$. We compute each of the powers $2^{2^j}$ with $j$ multiplications and it is easy to see, using Lemma 1, that we obtain

$$\widetilde{2^{2^j}} = 2^{2^j}(1+\theta_{2^j-1}).$$

Continuing using this lemma, we compute $2^e$ using (20) (multiply the powers with smaller exponent first) and obtain

$$\widetilde{2^e} = 2^e(1+\theta_{2^{k+1}-2}).$$

An extra multiplication with $\widetilde{m}$ yields

$$\widetilde{y_v} = y_v(1+\theta_{2^{k+1}+2k+2}) = y_v(1+\theta_{2^{k+2}})$$

as claimed. Assume finally that $2^k < \frac{\varepsilon}{2}$. Then, $\mathsf{u_{mach}} \le 2^{-2k} < \frac{\varepsilon^2}{4}$ and

$$|\theta_{2^{k+2}}| \;\le\; \frac{2^{k+2}\mathsf{u_{mach}}}{1-2^{k+2}\mathsf{u_{mach}}} \;\le\; \frac{2^k\varepsilon^2}{1-2^k\varepsilon^2} \;\le\; \frac{\varepsilon}{2-\varepsilon} \;\le\; \varepsilon$$

which finishes the proof. □

### 6.3 $\mathrm{NP}^{\mathsf{B}}_{\mathsf{ro}} \subset \mathrm{EXP}_{\mathsf{ro}}$

We can now show a key membership result for $\mathrm{EXP}^{[\mathsf{P}]}_{\mathsf{ro}}$.

**Theorem 4** *We have* $\mathsf{CircBFeas} \in \mathrm{EXP}^{[\mathsf{P}]}_{\mathsf{ro}}$.

PROOF. The general idea is to evaluate the circuit on the points of a canonical grid.

We consider the machine $\mathcal{M}$ given by the following code (here $n$ is the number of input gates of the circuit $\mathcal{C}$):

```
input 𝒞
compute n
set k := ⌊k_mach / 2⌋
for all v̄ ∈ {0,1}^{n(2k+2)} do
     compute y_v (overlined)
     evaluate f_𝒞(y_v)
accept if for one of these evaluations we obtain f_𝒞(y) ≥ 0
```

The arithmetic cost of $\mathcal{M}$ is easily bounded. Since $F_k$ contains $2^{\mathcal{O}(\mathsf{k}_{\mathsf{mach}})}$ numbers the grid $\mathscr{G}_k$ contains $2^{\mathcal{O}(n\mathsf{k}_{\mathsf{mach}})}$ points. To produce each of these points takes $\mathcal{O}(n\mathsf{k}_{\mathsf{mach}})$ arithmetic operations (Proposition 18), and to evaluate $\mathcal{C}$ at each of them an additional $\mathcal{O}(\mathsf{length}(\mathcal{C}))$ operations. It follows that the arithmetic cost of $\mathcal{M}$ is bounded by $2^{\mathcal{O}(\mathsf{length}(\mathcal{C})\mathsf{k}_{\mathsf{mach}})}(\mathsf{length}(\mathcal{C})\mathsf{k}_{\mathsf{mach}})^{\mathcal{O}(1)}$, a bound we can write in the form $2^{\mathcal{O}(\mathsf{length}(\mathcal{C})\mathsf{k}_{\mathsf{mach}})}$, as we wanted.

To prove the bound on the precision needed, we assume that $\mu_{\mathsf{Bfeas}}(\mathcal{C}) < \infty$ (and hence, that the same holds for $\mathsf{size}(\mathcal{C})$) and take $\mathsf{u}_{\mathsf{mach}} := \frac{1}{16(\mu_{\mathsf{Bfeas}}(\mathcal{C}))^2}$. We want to see that when $M$ works with this precision it correctly decides whether its input $\mathcal{C}$ is in $\mathsf{CircBFeas}$. Note that for $\mathsf{k}_{\mathsf{mach}}$ as above the unit roundoff of $F_k$ satisfies $u_k = \frac{1}{4\mu_{\mathsf{Bfeas}}(\mathcal{C})}$.

We divide by cases.

Assume first that $\mathcal{C}$ is feasible. In this case there exist points $x \in S_{\mathcal{C}}$. Let $x^*$ be one such point satisfying

$$\varrho_{\mathsf{Bfeas}}(\mathcal{C}) = \varrho_{\mathsf{eval}}(\mathcal{C}, x^*)2^{-\mathsf{mgt}(x^*)} > 0. \tag{21}$$

This implies that $2^{\mathsf{mgt}(x^*)} < \mu_{\mathsf{Bfeas}}(\mathcal{C}) = \frac{1}{4\sqrt{\mathsf{u}_{\mathsf{mach}}}}$ and therefore, that $\mathsf{mgt}(x^*) < \frac{1}{2}\mathsf{k}_{\mathsf{mach}} - 2 < k$. That is, $x^*$ belongs to $\mathsf{Range}(F_k)$.

This implies the existence of $\bar{v} \in \{0,1\}^{n(2k+2)}$ such that the corresponding $\overline{y_v} = (y_1, \ldots, y_n) \in \mathscr{G}_k$ satisfies, for $i = 1, \ldots, n$,

$$y_i = x_i^*(1+\delta) \quad \text{with } |\delta| < u_k = \frac{1}{4\mu_{\mathsf{Bfeas}}(\mathcal{C})}. \tag{22}$$

Since $\mathsf{u}_{\mathsf{mach}} \leq 2^{-2k}$ and $2^k \leq \frac{1}{4(\mu_{\mathsf{Bfeas}}(\mathcal{C}))}$, Proposition 18 ensures that the computation of $\overline{y_v}$ done by $\mathcal{M}$ returns a point $\widetilde{y}$ satisfying

$$\widetilde{y_i} = y_i(1+\delta) \qquad \text{with } |\delta| \leq \frac{1}{2\mu_{\mathsf{Bfeas}}(\mathcal{C})}. \tag{23}$$

From this inequality, together with (22), we deduce that

$$\widetilde{y_i} = x_i^*(1+\delta) \qquad \text{with } |\delta| \leq \varrho_{\mathsf{Bfeas}}(\mathcal{C}) < \varrho_{\mathsf{eval}}(\mathcal{C}, x^*) \tag{24}$$

the last inequality by (21).

Consider the computation of $\mathcal{M}$ corresponding to the point $\bar{v}$. It first produces the approximation $\widetilde{y}$ of $\overline{y_v}$ and then evaluates $f_{\mathcal{C}}(\widetilde{y})$ with a $\mathsf{u}_{\mathsf{mach}}$-computation $\gamma$.

We can associate to this computation the evaluation of $\mathcal{C}$ at $x^*$ that first approximates $x^*$ by $\widetilde{y}$ and then approximates $f_{\mathcal{C}}(\widetilde{y})$ with $\gamma$. We claim that this is a $\varrho_{\mathsf{eval}}(\mathcal{C}, x^*)$-evaluation. Indeed, the relative errors in the first process are bounded by $\varrho_{\mathsf{eval}}(\mathcal{C}, x^*)$ (because of (24)) and the second process is a $\varrho_{\mathsf{eval}}(\mathcal{C}, x^*)$-evaluation of $\mathcal{C}$ at $\widetilde{y}$ (we use $\mathsf{u_{mach}} < \varrho_{\mathsf{eval}}(\mathcal{C}, x^*)$ and Proposition 4). Putting these bounds together the claim follows.

The definition of $\varrho_{\mathsf{eval}}(\mathcal{C}, x^*)$ implies that this evaluation returns $f_{\mathcal{C}}(x^*) \geq 0$. But this implies that the $\mathsf{u_{mach}}$-computation of $\mathcal{M}$ above also yields $f_{\mathcal{C}}(x^*) \geq 0$. Which in turn implies that $\mathcal{M}$ accepts $\mathcal{C}$.

Assume now that $\mathcal{C}$ is not feasible. In this case, for all $x \in \mathbb{R}^n$, $\varrho_{\mathsf{eval}}(\mathcal{C}, x) \geq \varrho_{\mathsf{Bfeas}}(\mathcal{C}) \geq \mathsf{u_{mach}}$. This implies that, for all $\bar{v} \in \{0,1\}^{n(2k+2)}$, the $\mathsf{u_{mach}}$-evaluation of $\mathcal{C}$ at the computed approximation $\widetilde{y}$ of $\overline{y_v}$ yields $f_{\mathcal{C}}(\widetilde{y}) < 0$. This implies that $\mathcal{M}$ rejects $\mathcal{C}$.

We can now conclude since

$$\mathsf{k_{mach}} = \left\lceil \log \frac{1}{\mathsf{u_{mach}}} \right\rceil = \lceil \log 16(\mu_{\mathsf{Bfeas}}(\mathcal{C}))^2 \rceil \leq 2\,\mathsf{size}(\mathcal{C}) + \mathcal{O}(1).$$

This linear bound is well within the polynomial growth required in the definition of $\mathrm{EXP}_{\mathsf{ro}}^{[\mathsf{P}]}$. □

Essential to the fact the largest precision needed is linear in $\mathsf{size}(\mathcal{C})$ is the circumstance that the values of $f_{\mathcal{C}}(y)$ are computed by $\mathcal{M}$ *independently*, for all the points $y$ in $\mathscr{G}_k$. This fact would also be central in the proof that $\mathrm{NP}_{\mathsf{ro}}^{\mathsf{B}}$ is actually included in the subclass $\mathrm{PAR}_{\mathsf{ro}}$ of $\mathrm{EXP}_{\mathsf{ro}}$ of problems decidable in variable-precision parallel polynomial time. But we don't deal with parallelism in this paper.

**Corollary 4** *We have* $\mathrm{NP}_{\mathsf{ro}}^{\mathsf{B}} \subset \mathrm{EXP}_{\mathsf{ro}}^{[\mathsf{P}]}$ *and the inclusion is strict.*

PROOF. The first statement readily follows from Proposition 16. The second follows from Proposition 15. □

**Open Question 3** A major open question in this section is whether $\mathrm{NP}_{\mathsf{ro}}^{\mathsf{U}}$ is included in $\mathrm{EXP}_{\mathsf{ro}}$ or, equivalently, whether $\mathsf{CircFeas}$ belongs to $\mathrm{EXP}_{\mathsf{ro}}$. We conjecture that this question has a ppositive answer.

**Open Question 4** The question above can be strengthened. We know that $\mathsf{CircFeas} \in \mathrm{NP}_{\mathsf{dir}}^{\mathsf{U}}$ and that it is actually complete in this class for $\mathsf{P}$-reductions. This raises the question of whether $\mathsf{CircFeas} \in \mathrm{EXP}_{\mathsf{dir}}$ (a membership that would imply $\mathrm{NP}_{\mathsf{dir}}^{\mathsf{U}} \subset \mathrm{EXP}_{\mathsf{dir}}$).

These last open questions are both related to a finite-precision analysis of existing algorithms. The feasibility of an algebraic circuit $\mathcal{C}$ reduces to the feasibility of a

Boolean combination of the form

$$\bigvee_{i=1}^{K} \bigwedge_{j=1}^{q_i} \varphi_{ij}(x) *_{ij} 0$$

where $*_{ij} \in \{\geq, >\}$, $q_i$ is polynomially bounded in $\mathsf{length}(\mathcal{C})$, $K$ is exponentially bounded in $\mathsf{length}(\mathcal{C})$, and $\varphi_{ij}$ are straight-line programs of length polynomially bounded (and hence degree exponentially bounded) in $\mathsf{length}(\mathcal{C})$. There exist a number of algorithms showing that the feasibility of such a system can be done, under the assumption of infinite precision, in time exponential in $\mathsf{length}(\mathcal{C})$ [33, 45]. Hence, an accuracy analysis proving correctness of any of these algorithms under a $\mathsf{k_{mach}}$ bounded by an exponential function of $\mathsf{length}(\mathcal{C})$ and $\mu_{\mathsf{feas}}(\mathcal{C})$ would imply membership of $\mathsf{CircFeas}$ to $\mathrm{NP}^{\mathsf{U}}_{\mathsf{dir}}$ (and hence the inclusions $\mathrm{NP}^{\mathsf{U}}_{\mathsf{dir}} \subset \mathrm{EXP}_{\mathsf{dir}}$ and $\mathrm{NP}^{\mathsf{U}}_{\mathsf{ro}} \subset \mathrm{EXP}_{\mathsf{ro}}$). But these finite-precision analyses are, as of today, not at hand.

# 7 Final Remarks

## 7.1 Average complexity

In Section 3 we defined decision problems as pairs of subsets and condition numbers and in doing so, we put essentially no requirement on what a condition number function is. Remark 5 in that section elaborated a little on this generality. We now elaborate more, focusing on a viewpoint that has accompanied the development of condition numbers practically since the origins of the notion of condition.

These origins can be traced back to the condition number $\kappa(A)$ (in Example 1) introduced by Turing [59] and von Neumann and Goldstine [60] to understand the loss of precision in the solution of systems of linear equations. For a system $Ax = b$ this is the number of correct figures in the entries of $A$ and $b$ minus the this number for the computed solution $\widetilde{x}$. It follows from (9) that this number is about $3\log n + \log\kappa(A)$. The fact, however, that $\kappa(A)$ is not known a priori (and computing it from $A$ is a process with the same shortcomings as those of solving $Ax = b$) prompted von Neumann and Goldstine to propose studying $\log\kappa(A)$ as a random variable. Their paper [61] exhibited some results in this direction assuming $A$ to be Gaussian (i.e., having its entries independent and normally distributed). The state of the art for this assumption was shown by Alan Edelman [27] who proved that, for Gaussian real or complex $n \times n$ matrices, we have

$$\mathbb{E}\log\kappa(A) = \log n + C + o(1) \tag{25}$$

where $C = 1.537$ in the real case and $C = 0.982$ in the complex. As a consequence, the loss of precision in the solution of $Ax = b$ is, on the average, of the order of $4\log n$. A different algorithm, analyzed in terms of a different condition number, would produce, for the same underlying probability measure a (likely) different dependence on $n$ and the comparison of these dependences translates into a comparison of the two algorithms' efficiencies.

In 1997, Steve Smale advocated studying the complexity of an algorithm in a similar manner. One would first derive complexity bounds in terms of input length and a condition number and then eliminate the latter by endowing the set of inputs of length $n$ with a probability measure (usually the standard Gaussian) and bounding the expected value of this condition number (or of its logarithm, if appropriate) by a function of $n$. One thus obtains *average complexity bounds* depending on the input length only. A convenient feature of this procedure is the fact that one is free to choose the condition number. Different analyses for a computational problem may rely on different condition numbers (sometimes for the same algorithm!) and algorithms may be compared by the average complexity bounds shown on these analyses. A case at hand is the polyhedral feasibility problem described in Example 5. We mentioned in this example that algorithmic solutions to this problem have been proposed whose analyses are based on a variety of condition numbers (in addition to $\mathscr{C}(A)$ defined there). Expected values for (the log of) these condition numbers allows for a comparison of the efficiency of these algorithmic solutions. We mention here that in [11] it is proved that for Gaussian matrices $A \in \mathbb{R}^{m\times n}$

$$\mathbb{E} \log \mathscr{C}(A) = 2\log(m+1) + 3.31. \tag{26}$$

This means that the contribution of the condition in the complexity bound mentioned in Example 5 (continued) is dominated, on the average, by that in terms of the dimension of $A$. The average complexity of the algorithm is $\mathcal{O}(n^{3.5}\log n)$. Similarly for the average of the required $\mathsf{k}_{\mathsf{mach}}$, which turns to be $\mathcal{O}(\log n)$.

Probabilistic analysis is therefore a way to reduce the arbitrariness in the choice of condition numbers. We won't go deeper into these ideas but point instead to the recent monograph [10], where condition numbers are the central character and their probabilistic analysis a recurrent theme.

### 7.2 Recursive analysis

We mentioned in §1.3 a stream of research that approaches numerical computations based on Turing machines. Detailed expositions of this research are in [38, 62]. It is worth to point out the differences with the theory we have developed in this paper. A first major difference is made clear early on [38, page 4] where the author mentions that, despite the fact that "the floating-point model is the usual model of computation in numerical analysis", it not the model used in the theory. Indeed, a look at this theory reveals it to be closer to fixed-point arithmetic. A second major difference is the absence of the notion of conditioning. Again, an examination of [38] reveals that this notion is absent in the book.

None of these differences is in detriment of an approach that has undeniably provided understanding of some numerical computations. But they both row against the agenda cited at the end of §1.1, namely "to incorporate round-off error [and] condition numbers into [the Blum-Shub-Smale] development."

Our theory attempts to stay close to the design and analysis of algorithms in numerical analysis (as described in [35]). It is also close to discrete computation in the measure that our finite-precision computations can be simulated by Turing machines. In these simulations real data, both machine constants and input, are replaced by floating-point numbers (with unrestricted exponents) and then operated. But these simulations do not preserve polynomial time because of the presence of unrestricted exponents (for every polynomial bound, just reading a real number with a sufficiently large magnitude will have a cost beyond this bound).

Our choices for machine model, size, and cost were taken, as wished in [6], "to bring machines over $\mathbb{R}$ closer to the subject of numerical analysis." We actually strived to bring them closer to the literature in numerical analysis (where unrestricted exponents dominate).

We nonetheless mention that a modification of our theory that makes it closer to the discrete framework is possible. One defines the size of $(u, x) \in \mathcal{I}$ by

$$\mathsf{size}(u, x) := \mathsf{length}(u, x) + \mathsf{mgt}(x) + \lceil \log_2 \boldsymbol{\mu}(u, x) \rceil.$$

Then, the basic version of polynomial time cost is obtained by writing $\mathsf{Arith}(\mathsf{length}(u, x) + \mathsf{mgt}(x), \mathsf{Prec}(\mathsf{size}(u, x)))$ in Definition 5. We have the feeling, though, that this theory would be more removed from numerical analysis' thinking. We have therefore not pursued this avenue.